\documentclass[twocolumn,english,aps,pra,reprint, superscriptaddress,showpacs,longbibliography,showkeys]{revtex4-2}
\usepackage{amsmath,amssymb,bbm,mathrsfs,bm,braket,color,graphicx,comment,xcolor,dsfont,multirow}
\usepackage{subcaption}
\usepackage{tikz}
\usepackage{amsmath}
\usetikzlibrary{decorations.pathreplacing}
 
\definecolor{tealline}  {RGB}{ 50,140,110}
\definecolor{purpborder}{RGB}{ 90,100,175}
\definecolor{purpdark}  {RGB}{ 70, 80,160}
\definecolor{darkorange}{RGB}{200,100,0}

\usepackage[colorlinks,citecolor=blue,urlcolor=blue,linkcolor = blue]{hyperref}
\usepackage[mathscr]{euscript}

\usepackage{multirow}
\usepackage[shortlabels]{enumitem}
\usepackage{physics}
\usepackage{tikz, ifthen}
\usetikzlibrary{arrows.meta}
\usetikzlibrary{decorations.pathreplacing,calc}
\usepackage{float}
\makeatletter
\let\newfloat\newfloat@ltx
\makeatother
\usepackage{algorithm}
\usepackage[noend]{algpseudocode}
\usepackage{orcidlink}

\begin{document}

\author{Robert~J.~Banks~\orcidlink{0009-0004-5198-1651}}
\thanks{These two authors contributed equally; \\
r.banks@parityqc.com \\
a.crippa@parityqc.com}
\affiliation{Parity Quantum Computing Germany GmbH, Schauenburgerstraße 6, 20095 Hamburg, Germany}

\author{Arianna~Crippa~\orcidlink{0000-0003-2376-5682}}
\thanks{These two authors contributed equally; \\
r.banks@parityqc.com \\
a.crippa@parityqc.com}
\affiliation{Parity Quantum Computing Germany GmbH, Schauenburgerstraße 6, 20095 Hamburg, Germany}

\author{Matthias~Traube~\orcidlink{0000-0002-4229-6829}}
\affiliation{Parity Quantum Computing Germany GmbH, Schauenburgerstraße 6, 20095 Hamburg, Germany}

\author{Josua~Unger~\orcidlink{0000-0003-1633-1843}}
\affiliation{Parity Quantum Computing GmbH, A-6020 Innsbruck, Austria}

\author{Christian~Ertler~\orcidlink{0000-0002-9820-1120}}
\affiliation{Parity Quantum Computing Germany GmbH, Schauenburgerstraße 6, 20095 Hamburg, Germany}

\author{Wolfgang~Lechner~\orcidlink{0000-0003-3662-1020}}
\affiliation{Parity Quantum Computing Germany GmbH, Schauenburgerstraße 6, 20095 Hamburg, Germany}
\affiliation{Parity Quantum Computing GmbH, A-6020 Innsbruck, Austria}
\affiliation{Institute for Theoretical Physics, University of Innsbruck, A-6020 Innsbruck, Austria}

\begin{abstract}
Parameterized Instantaneous Quantum Polynomial (IQP) circuits have proven useful in quantum generative learning models, particularly for binary distributions. However, when applied to non-binary datasets, they exhibit notable limitations: mapping integer values into qubit-compatible binary representations often destroys the original metric structure of the data.
In this paper we aim to extend them to a \textit{qudits} formulation operating on an integer mapping of the data. The IQP quantum circuit is adapted to encode each integer valued pixel into a bit-string of fixed length and quantum gates are transformed to follow the qudit formalism.
As a generative machine learning approach, a suitable loss function for the circuit training and the calculation of the covariance matrix among features are developed and validated on the energy deposits from single-particle electron showers in the electromagnetic calorimeter of the CLIC detector. 
The method proposed in this work can be also extended to other applications that utilize quantum generative machine learning for non-binary data.
\end{abstract}

\date{\today}

\title{Qudit extension of parameterized IQP circuits:\\ A generative quantum machine learning approach to integer data}

\maketitle

\section{Introduction}
\label{sec:intro}

Generative machine learning methods seek to approximate the underlying probability distribution of a dataset, enabling the synthesis of new samples that faithfully reflect the statistical structure of the original data. 
Quantum systems are particularly well suited to this task: measurement outcomes are governed by the Born rule, providing a physically motivated and hardware-native source of stochasticity, while the exponential dimensionality of the Hilbert space allows complex distributions to be represented compactly, and sampling is efficiently achieved~\cite{perdomo2018opportunities,Liu_2018,benedetti2019generative,coyle2020born}.
Currently, quantum generative models present limitations, such as the presence of barren plateaus~\cite{mcclean2018barren,cerezo2021cost,ortiz2021entanglement,rudolph2024trainability}.
In Recio-Armengol \textit{et al.}~\cite{recio2025train}, the authors show a possible path to scaling  generative quantum machine learning by employing parameterized instances of instantaneous quantum polynomial (IQP) circuits~\cite{nakata2014diagonal,bremner2016average}.
The key idea is to avoid directly sampling during training and instead learn the circuit parameters using classical optimization methods. Training is based on efficiently computing expectation values or moments. As a consequence, the training can be carried out entirely on classical hardware~\cite{Nest_2010} by optimizing a statistical discrepancy between the model and the target data distribution. A central component of this optimization is the Maximum Mean Discrepancy (MMD)~\cite{gretton2012kernel}, a kernel-based test statistic that measures whether two distributions are identical. Ref.~\cite{rudolph2024trainability} shows that the MMD can be rewritten as a combination of expectation values of Pauli-$Z$ operators.
The trained IQP circuit is then executed on a quantum device to generate samples from the modeled distribution. The motivation is that sampling from the trained IQP circuit is believed to be classically hard in general, thus requiring quantum hardware~\cite{bremner2011classical,bremner2016average, marshall2024improved,hangleiter2023computational}. Accordingly, IQP circuits provide a good candidate for investigating trainability and scalability using current classical resources, while preserving the prospect of achieving quantum advantage in the future. 
Within the IQP circuits framework, the dataset considered has mainly been binary (or reduced to binary) to be able to fit into a qubit distribution. 

In this work, the IQP circuit analysis from Ref.~\cite{recio2025train} is extended to be able to address non-binary data distributions.
In particular, a \textit{qudit}-based (or \textit{integer}-based) framework for IQP circuits is presented. 
The goal is to develop a methodology that avoids the need to binarize the target data distribution. Instead, the approach operates on an integer mapping of the data. We describe in this paper all the mathematical tools required to calculate the expectation values of diagonal operators in the qudit formalism. We furthermore modify the MMD loss function to address a cyclic distance between qudit states.

Our method is validated via training of the circuit on the calorimeter electron showers energy deposits.
Simulating electromagnetic particle showers is a fundamental task in high energy physics (HEP). 
In HEP experiments, detectors function as imaging systems that record the decay products of particle-collision events. Calorimeters are a central component of these detectors: as a high-energy particle traverses the dense material, it deposits energy and initiates a cascade of secondary particles. This energy is measured by detector cells, producing images that encode signatures of the primary particle’s type and energy.
Modeling this distribution is challenging because it requires capturing correlations in the energy deposition structure, and classical simulations based on Monte Carlo methods are computationally intensive. Quantum computing may offer a potential alternative framework for addressing these computational demands.
Quantum generative models could also serve as simulators, with applications in tasks such as data reconstruction and classification in high energy physics.
Our target distribution is the $z$-axis projection of the $51 \times 51 \times 25$ cell window in the electromagnetic calorimeter energy (ECAL) of Ref.~\cite{belayneh2020calorimetry}. Each of the 3D images show incoming particles entering from $z = 0$, at the center of the $(x, y)$ transverse plane $(x = y = 25)$. The publicly available dataset~\cite{pierini2020clic} contains the energy deposits from single-particle electron showers in the ECAL calorimeter of the CLIC~\cite{linssen2012physics} detector at a fixed angle ($\theta=\pi/2$).

While we have applied our method to HEP data, recent works considered quantum generative methods to generate other datasets, for instance SAT4 satellite images explored in Ref.~\cite{liepelt2026exponentialcapacityscalingclassical} (based on generative adversarial networks~\cite{Dallaire_Demers_2018, Zoufal_2019}).

\vspace{0.2cm}

The paper is structured as follows: Sec.~\ref{sec:framework} describes the numerical methods and mathematical formalism for the generative approach with IQP circuits for qudits.
This section begins with describing the quantum generative learning method in Sec.~\ref{subsec:quant_gen_learning}. After which we discuss a possible framework to generalize IQP circuits for qubits~\cite{armengol2026iqpoptfastoptimizationinstantaneous} (Sec.~\ref{subsec:binary_iqp}), into IQP circuits for qudits (Sec.~\ref{subsec:int_iqp}), focusing on: how the quantum circuit can be adapted to integer data, ansatz design, classically estimating expectation values, and suitable loss functions for the training of the model. The model is  summarized in Sec.~\ref{sec:model_summary}.
Sec.~\ref{sec:results} applies the aforementioned method to the energy deposit dataset of the CLIC calorimeter. In particular, we describe: how the dataset resolution is reduced (Sec.~\ref{sec:data_reduction}), and the training of the IQP circuit in Sec.~\ref{sec:iqp_training} for two case studies with 6 and 12 pixels.
Sec.~\ref{sec:circ_validation} validates the results of the previous section, by computing the covariance and correlation matrices between the generated data and data withheld from the training.
Sec.~\ref{sec:circ_implement} presents a possible strategy to implement the
IQP circuit on quantum hardware, using either the Parity Twine method \cite{dreier2025connectivityawaresynthesisquantumalgorithms} or mid-circuit measurements.
Finally, Sec.~\ref{sec:discussion} summarizes the main findings and provides an outlook on future directions and potential further investigations.

In App.~\ref{app:decomp} qudit-based operators and related identities used in the development of the framework is provided for ease of reference. The mathematical details of the loss function are described in App.~\ref{app:loss_function}. In App.~\ref{app:potts} the inclusion of symmetries in the IQP ansatz is explored with a test on the Potts model in App.~\ref{app:potts_res}. 
In App.~\ref{app:add_results} preliminary results on the 25 pixels HEP images dataset are reported.
App.~\ref{app:qml_works} gives a summary on prior quantum generative approaches on HEP calorimeter images.

During the preparation of this manuscript, another proposal of an extension of IQP circuits for generating calorimeter images~\cite{slim2026iqpbornmachinecalorimeter} was published. A brief discussion on this approach can be found in App.~\ref{app:iqp_sparse}. Shortly after the upload of the manuscript, Huang \textit{et al.}~\cite{huang2026spectralbornmachinesclassically} carried out a work which generalizes IQP Born machines in a similar manner.

\section{Extension of IQP circuits to integer data}
\label{sec:framework}

This section outlines the extension of parameterized IQP circuits from binary data to integer data. In order to do this, we give a high-level summary of the aims of quantum generative-learning in Sec.~\ref{subsec:quant_gen_learning} and the established binary parameterized IQP case in Sec.~\ref{subsec:binary_iqp}. From there, Sec.~\ref{subsec:int_iqp} discusses how such parameterized IQP circuits can be better tailored to integer data, while maintaining the ability to train classically.

\subsection{Quantum generative learning}
\label{subsec:quant_gen_learning}
Consider a \textit{training dataset} of vectors $\mathbf{x}$, sampled from a target distribution $p(\mathbf{x})$. In this work we primarily focus on the case of binary vectors $\mathbf{x} \in \{0,1\}^n$ and integer vectors $\mathbf{x} \in \{0,..,d-1\}^n$, with $d-1$ being the maximum possible integer. In both cases $n$ is the number of pixels (or features) in the training dataset. The training set can be used to construct an empirical probability distribution $\tilde{p}(\mathbf{x})$.
A generative model is a parameterized conditional distribution $q_{\boldsymbol{\theta }}(\mathbf{x})$ where $\boldsymbol{\theta }$ is a vector of trainable parameters.
The aim of quantum generative learning is to find the optimal parameters $\boldsymbol{\theta }^*$ such that samples drawn from $q_{\boldsymbol{\theta }}$ closely resemble those of $p$, thus enabling the generation of new and previously unseen data. 
The process of training requires a loss function $\mathcal{L}(\boldsymbol{\theta })$ which estimates the distance between the model $q_{\boldsymbol{\theta }}(\mathbf{x})$ and the training distribution $\tilde{p}(\mathbf{x})$. In the case of perfect learning of the empirical probability distribution, $\mathcal{L}(\boldsymbol{\theta })=0$ if and only if $q_{\boldsymbol{\theta }}(\mathbf{x})=\tilde{p}(\mathbf{x})$. The model is \textit{overfitting}, i.e. perfectly learning $\tilde{p}(\mathbf{x})$ but not the true target $p(\mathbf{x})$. As a consequence, the aim of the generative model is to reduce the training loss but not perfectly minimize it.

\subsection{The binary formulation}
\label{subsec:binary_iqp}

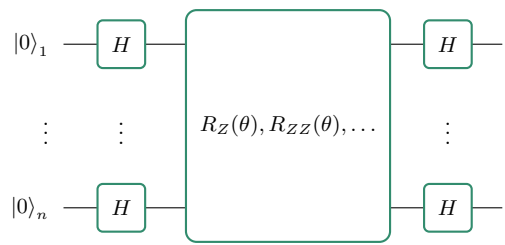
\begin{figure}[htp]
    \centering
\begin{tikzpicture}[scale=0.9, every node/.style={scale=0.9}]
 
  \def\ya{0}
  \def\ymid{-1.2}
  \def\yd{-2.4}
 
  \def\xKr{1.05}
  \def\xW{1.15}
  \def\xHl{2.0}
  \def\xBl{2.95}
  \def\xBr{5.95}
  \def\xHr{6.80}
  \def\xZ{7.60}
 
  \foreach \y in {\ya,\yd}{\draw (\xW,\y)--(\xZ,\y);}
 
  \node[font=\small,anchor=east] at (\xKr,\ya){$\ket{0}_1$};
  \node[font=\small,anchor=east] at (\xKr,\ymid){$\vdots$};
  \node[font=\small,anchor=east] at (\xKr,\yd){$\ket{0}_n$};
 
  \node[font=\small] at (\xHl,\ymid){$\vdots$};
  \node[font=\small] at (\xHr,\ymid){$\vdots$};
 
  \draw[tealline,thick,fill=white,rounded corners=4pt]
    (\xBl,\yd-0.5) rectangle (\xBr,\ya+0.5);
  \node[font=\small] at ({(\xBl+\xBr)/2},{(\ya+\yd)/2})
    {$R_Z(\theta),R_{ZZ}(\theta),\ldots$};
 
  \foreach \y in {\ya,\yd}{
    \draw[tealline,thick,fill=white,rounded corners=2pt]
      (\xHl-0.35,\y-0.35) rectangle (\xHl+0.35,\y+0.35);
    \node[font=\small] at (\xHl,\y){$H$};
    \draw[tealline,thick,fill=white,rounded corners=2pt]
      (\xHr-0.35,\y-0.35) rectangle (\xHr+0.35,\y+0.35);
    \node[font=\small] at (\xHr,\y){$H$};
  }
 
\end{tikzpicture}

\caption{\textbf{IQP quantum circuits structure for the qubit implementation:} Each qubit is initialized in the zero state $\ket{0}$. A layer of Hadamards is followed by a sequence of diagonal multi-body Pauli-$Z$ rotations. A final layer of Hadamard gates is then applied.}
    \label{fig:q_circuits_qubit}
\end{figure}

Recio-Armengol \textit{et al.}~\cite{recio2025train} explored training quantum generative models classically with the generation to be realized on real quantum hardware. This requires sample generation to be classically hard, while the loss function remains efficiently computable on classical hardware. To satisfy these conditions, they utilize parameterized IQP circuits. Sampling from IQP circuits is expected to be hard for classical algorithms~\cite{bremner2011classical,bremner2016average, marshall2024improved}, while estimating expectation values remains classically tractable~\cite{Nest_2010}. For the parameterized IQP circuits each computational basis state encodes a possible outcome of the generative model. Therefore, sampling is required for data generation in the deployment of the model. 
The aim is to learn the amplitudes of the state-vector, so that the amplitudes  correspond to the distribution to be learned. To do this it is necessary to choose a loss function. Recio-Armengol \textit{et al.}\ take the loss function to be the Maximum Mean Discrepancy with a Gaussian kernel. It was shown by Rudolph \textit{et al.}\ \cite{rudolph2024trainability} that the MMD in this case can be written as a probabilistic sum over expectation values. This allowed Recio-Armengol \textit{et al.}\ to train classically in regimes where direct classical simulation is intractable. 
Note that the use of the MMD to train (classical) generative models~\cite{li2015generative} was extended to the quantum setting in Ref.~\cite{Liu_2018}.

A parameterized IQP circuit on $n$ binary features  (and equivalently $n$ qubits) is a circuit comprised of an initial layer of Hadamard gates on the all-zero state $\ket{0}^{\otimes n}$, followed by a set of diagonal $k$-body Pauli-$Z$ rotations (typically one- and two-body). A final layer of Hadamards is then applied. The circuit is illustrated in Fig.~\ref{fig:q_circuits_qubit}. The full parameterized unitary is
\begin{equation}
    \label{eq:bin_iqp_u}
    U(\boldsymbol{\theta }) = \prod_{j=1}^{N_\text{gates}} \exp(i\theta_j X_{\boldsymbol{g}_j}),  
\end{equation}
with $ \boldsymbol{\theta}$ the vector of trainable parameter and $\boldsymbol{g}_j$ a binary vector of length $n$ that encodes where each gate acts non-trivially. The operator $X_{\boldsymbol{g}_j}$ is therefore short-hand for:
\begin{equation}
    X_{\boldsymbol{g}_j} = \bigotimes_{i=1}^n X^{{g_j}_i},
\end{equation}
where ${g_j}_i$ denotes the $i^\text{th}$ component of $\boldsymbol{g}_j$. Since all generators in the exponent commute in Eq.~\eqref{eq:bin_iqp_u}, the circuit is invariant under the ordering of the individual gate operations.

\subsection{Adapting IQP circuits to the integer case}
\label{subsec:int_iqp}

Existing approaches to IQP circuits~\cite{recio2025train,lerch2026iqp} are mainly designed for datasets that are inherently binary. To adapt these methods for the generation of energy spectra, it is important to address the limitations of binary encoding. When integers are mapped to a qubit-compatible binary format, the metric structure of the original data is often lost. Specifically, the Hamming distance between binary strings does not align with the Euclidean distance of the original values; for example, the values `0' and `8' are numerically far apart, yet their binary representations (\texttt{0000} and \texttt{1000}) are only one bit apart. Consequently, a standard loss function on binary strings is expected to fail to capture the true underlying distance between these points.

\subsubsection{Adapting the circuit}

The current binary IQP formulation consists of a trainable circuit diagonal in the computational basis, with a layer of Hadamard gates on each side as shown in Fig.~\ref{fig:q_circuits_qubit}. In this paper the circuit is adapted to the case where each integer has been encoded into a bit-string of length $b$. Therefore, one can think of the circuit as consisting of $n$ qudits (each qudit encoding one feature), each of dimension $d=2^b$. 

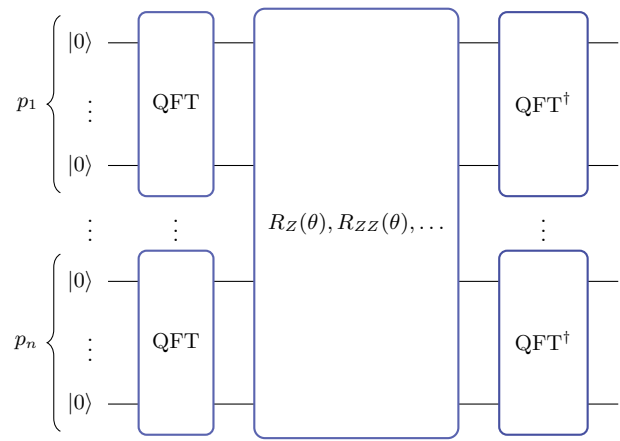
\begin{figure}[htp]
    \centering
\begin{tikzpicture}[scale=0.9, every node/.style={scale=0.9}]
 
  \def\yat{0}
  \def\yab{-1.8}
  \def\ygap{-2.65}
  \def\ybt{-3.5}
  \def\ybb{-5.3}
 
  \def\xKr{1.05}
  \def\xW{1.15}
  \def\xQl{2.15}   
  \def\xBl{3.30}
  \def\xBr{6.30}
  \def\xQr{7.55}   
  \def\xZ{8.65}
  \def\xBrace{0.45}
 
  \foreach \y in {\yat,\yab,\ybt,\ybb}{\draw (\xW,\y)--(\xZ,\y);}
 
  \node[font=\small,anchor=east] at (\xKr,\yat){$\ket{0}$};
  \node[font=\small,anchor=east] at (\xKr,\yab){$\ket{0}$};
  \node[font=\small,anchor=east] at (\xKr,{\yat+(\yab-\yat)/2}){$\vdots$};
 
  \node[font=\small,anchor=east] at (\xKr,\ybt){$\ket{0}$};
  \node[font=\small,anchor=east] at (\xKr,\ybb){$\ket{0}$};
  \node[font=\small,anchor=east] at (\xKr,{\ybt+(\ybb-\ybt)/2}){$\vdots$};
 
  \node[font=\small,anchor=east] at (\xKr,\ygap){$\vdots$};
  \node[font=\small] at (\xQl,\ygap){$\vdots$};
  \node[font=\small] at (\xQr,\ygap){$\vdots$};
 
  \draw[decorate,decoration={brace,amplitude=5pt}]
    (\xBrace,\yab-0.4)--(\xBrace,\yat+0.4) node[midway,left=6pt]{$p_1$};
  \draw[decorate,decoration={brace,amplitude=5pt}]
    (\xBrace,\ybb-0.4)--(\xBrace,\ybt+0.4) node[midway,left=6pt]{$p_n$};
 
  \draw[purpborder,thick,fill=white,rounded corners=4pt]
    (\xBl,\ybb-0.5) rectangle (\xBr,\yat+0.5);
  \node[font=\small] at ({(\xBl+\xBr)/2},{(\yat+\ybb)/2})
    {$R_Z(\theta),R_{ZZ}(\theta),\ldots$};
 
  \foreach \ytop/\ybot in {\yat/\yab, \ybt/\ybb}{
    \draw[purpborder,thick,fill=white,rounded corners=3pt]
      (\xQl-0.55,\ybot-0.45) rectangle (\xQl+0.55,\ytop+0.45);
    \node[font=\small] at (\xQl,{(\ytop+\ybot)/2}){$\mathrm{QFT}$};
    \draw[purpdark,thick,fill=white,rounded corners=3pt]
      (\xQr-0.65,\ybot-0.45) rectangle (\xQr+0.65,\ytop+0.45);
    \node[font=\small] at (\xQr,{(\ytop+\ybot)/2}){$\mathrm{QFT}^\dagger$};
  }
\end{tikzpicture}
\caption{\textbf{IQP quantum circuits structure for the qudit implementation:} For each of the $n$ pixels, $\log_2 d $ qubits (with $d$ restricted to being a power of 2) are used. These qubits are initialized in the zero state $\ket{0}$. A layer of QFTs is followed by a sequence of multi-body diagonal rotations. A final layer of inverse QFT gates is then applied.}
    \label{fig:q_circuits_qudit}
\end{figure}

The Hadamard operator for a qudit is denoted as~\cite{pudda2024generalised},
\begin{equation}
    \label{eq:qft}
    H_d = \frac{1}{\sqrt{d}}\sum_{j,k=0}^{d-1}  \omega_d^{jk} \ket{j}\bra{k},
\end{equation}
with $\omega_d = e^{2\pi i/d}$ being the $d^\text{th}$ root-of-unity.  Eq.~\eqref{eq:qft} corresponds to the standard Quantum Fourier Transform (QFT) for $d=2^b$.  The layer of Hadamards is therefore replaced with a QFT, acting separately on each qudit. This is illustrated in Fig.~\ref{fig:q_circuits_qudit}. The trainable part of the circuit is left as being diagonal in the computational basis and is discussed further in Sec.~\ref{subsubsec:ansatz}.

Transforming the circuit means that expectation values cannot be estimated using the same expression as in the binary IQP case. However, we wish to still be able to train classically. The next section details how certain expectation values can be estimated.

\subsubsection{Estimating expectation values}

Having looked at the generalization of the Hadamard to qudits, we now define the Pauli-$Z$ and Pauli-$X$ operators to be~\cite{pudda2024generalised}:
\begin{align}
    Z_d &= \sum_{j=0}^{d-1} \omega_d^j \ket{j}\bra{j},\label{eq:Zd}\\
    X_d &= \sum_{j=0}^{d-1}  \ket{j+1 \bmod d}\bra{j},\label{eq:Xd}
\end{align}
with the latter acting as a cyclic shift. Note that for qudits, these Pauli operators are defined such that applying them $d$ times returns the identity operator, $X_d^d = Z_d^d = I_d$. These operators are unitary but not Hermitian. From now on the explicit index $d$ is dropped, it should be tacitly assumed (unless otherwise specified) that the qudits have dimension $d$.

This section describes how to classically estimate expectation values of powers of $Z$ for the parameterized qudit IQP circuits, with $n$ qudits. 
The parameterized diagonal part of the ansatz is taken to be:
\begin{equation}
    \label{eq:u_diag}
    U_{\text{diag}} = 
\exp \left( i \sum_{j}\theta_j\, Z_{\mathbf g_j} \right),
\end{equation}
where $\mathbf g_j \in \{0,..,d-1\}^{(n)}$ denotes the power of $Z$ acting on each qudit. That is to say,
\begin{equation}
    Z_{\mathbf g_j} = \bigotimes_{k=1}^n Z^{{g_j}_k},
\end{equation}
where ${g_j}_k$ denotes the $k^{\text{th}}$ element of $\mathbf g_j$. The parameters $\theta_j$ are generally complex, and cannot necessarily be trained independently, however any diagonal ansatz can be expressed in the form of Eq.~\eqref{eq:u_diag}.

The expectation of $Z_{\mathbf a}$, where $\mathbf a \in \{0,..,d-1\}^{(n)}$ denotes the power of $Z$ being measured on each qudit, can be stochastically estimated:
\begin{align}
     \langle Z_{\boldsymbol{a}} \rangle
    &= \bra{0} H^{\otimes n, \dagger} U_{\text{diag}}^\dagger H^{\otimes n} Z_{\boldsymbol{a}}  H^{\otimes n, \dagger} U_{\text{diag}} H^{\otimes n} \ket{0} \nonumber \\
    & = \frac{1}{d^n} \sum_{\boldsymbol{z'},\boldsymbol{z}} \bra{\boldsymbol{z'}} U_{\text{diag}}^\dagger X_{\boldsymbol{d}-\boldsymbol{a}}  U_{\text{diag}} \ket{\boldsymbol{z}} \nonumber \\
    & =\frac{1}{d^n} \sum_{\boldsymbol{z'},\boldsymbol{z}} e^{-i \sum_{j}\theta_j^*\, \omega^{-\mathbf{g_j}\cdot \boldsymbol{z'}}  } e^{i \sum_{j}\theta_j^*\, \omega^{\mathbf{g_j}\cdot \boldsymbol{z}}  } \bra{\boldsymbol{z'}} X_{\boldsymbol{d}-\boldsymbol{a}} \ket{\boldsymbol{z}} \nonumber \\
    \label{eq:z_exp}
    & = \mathbb{E}_{\boldsymbol{z}\sim U} \exp \left[i \sum_j \left(\theta_j \omega^{\mathbf{g_j}\cdot \boldsymbol{z}} - \theta_j^*\omega^{-\mathbf{g_j} (\boldsymbol{z}-\mathbf{a})}\right) \right],
\end{align}
where $\boldsymbol{z}\sim U$ denotes sampling $\boldsymbol{z}$ uniformly from $\{0,..,d-1\}^{(n)}$. Monte Carlo sampling is then used, as in Ref.~\cite{recio2025train}, to efficiently estimate  $\langle Z_{\boldsymbol{a}} \rangle$.

Since the expectation values of the form $\langle Z_{\boldsymbol{a}} \rangle$ can be estimated for the qudit IQP circuits, it is necessary to decompose any observable of interest into a linear combination of $\langle Z_{\boldsymbol{a}} \rangle$s. A list of the decompositions used in this paper can be found in App.~\ref{app:decomp}.

\subsubsection{Selecting a loss-function}

The loss-function used in Ref.~\cite{recio2025train} is the Maximum Mean Discrepancy:
\begin{equation}
\mathrm{MMD}^2(p, q_\theta)
= \mathbb{E}_{\mathbf{a} \sim P_\sigma(\mathbf{a})}
\left[\left( \langle Z_\mathbf{a} \rangle_p - \langle Z_\mathbf{a} \rangle_{q_\theta} \right)^2\right]
\end{equation}
for qubits in the binary IQP case, where $\mathbf{a} \sim P_\sigma(\mathbf{a})$ denotes sampling each term in $\mathbf{a}$ from an independent and identical Bernoulli trial. The success probability of the Bernoulli trial is parameterized by the bandwidth $\sigma$. The subscripts of the expectation values denote over which distribution the expectation value is taken ($p$ is the target distribution, $q_\theta$ is the generated distribution). 
We extend this straightforwardly to the qudit case:
\begin{equation}
    \text{MMD}^2(p, q_\theta)
= \mathbb{E}_{\mathbf{a} \sim P_\sigma(\mathbf{a})}
\left[\abs{ \langle Z_\mathbf{a} \rangle_p - \langle Z_\mathbf{a} \rangle_{q_\theta}}^2\right],
\end{equation}
where $P_\sigma$ is a probability density function with non-zero support on all $\mathbf{a}$, $\langle Z_\mathbf{a} \rangle_{q_\theta}$ follows Eq.~\eqref{eq:z_exp} and 
\begin{equation}
\left\langle Z_{\boldsymbol a} \right\rangle_{p}
=\mathbb{E}_{\boldsymbol{x}\sim p} \,\omega^{\mathbf{a}\cdot \boldsymbol{x}} .
\end{equation}

This loss function is real and greater than or equal to zero (it is zero if and only if $p=q_\theta$).
Inspired by a Gaussian Kernel MMD, in Sec.~\ref{sec:iqp_training} we take each coefficient of $\boldsymbol{a}$ to be sampled from an identical and independent distribution given by:
\begin{align}\label{eq:w_m}
&p_{\sigma}(a) =
\frac{\tilde{w}_{\sigma}(a)}{\sum_{k=0}^{d-1} \tilde{w}_{\sigma}(a)},
\\
& \text{with} \ \tilde{w}_{\sigma}(a)=
\exp\!\left(-2\pi^{2}\sigma^{2}\frac{\min^{2}(a,d-a)}{d^{2}}\right)\nonumber.
\end{align}
where $\sigma$ controls the width of the distribution and $a\in \{0,...,d-1\}$. In App.~\ref{app:loss_function} more details on the motivation behind this specific form of $p_{\sigma}(a)$ are provided. Prior work on binary IQP circuits have explored deciding the form of $p_{\sigma}(\mathbf{a})$ as a trainable part of a hybrid machine-learning model \cite{kurkin2025universality}.

\subsubsection{Ansatz design}
\label{subsubsec:ansatz}

The IQP circuit contains a parameterized unitary which is diagonal in the computational basis. The form of this diagonal ansatz is left as a design choice. In Sec.~\ref{sec:results} a hardware efficient ansatz~\cite{Leone_2024} is considered, on the assumption that the hardware is ultimately qubit based, all one- and two- body qubit $Z$ rotations available on the hardware are utilized. In App.~\ref{app:potts} an alternative ansatz informed by the problem at hand is explored. For the ansatz to be evaluated using Eq.~\eqref{eq:z_exp}, the ansatz needs to be expressed in the form of Eq.~\eqref{eq:u_diag}. Let us consider an ansatz of the form
\begin{equation}
    U_\text{diag} = e^{i H_\text{diag}},
\end{equation}
with
\begin{equation}
    H_\text{diag} = \sum_{i=1}^{N_\text{params}} \alpha_i \bigotimes_{j=1}^n D_j^{(i)},
\end{equation}
where $\alpha_i$ are real trainable parameters and $D_j^{(i)}$ diagonal Hermitian operators acting on the $j^\text{th}$ qudit. There are $N_\text{params}$ trainable parameters in total. The diagonal operators can be expressed as a linear combination of $Z$,
\begin{equation}
    H_\text{diag} = \sum_{i=1}^{N_\text{params}} \alpha_i \bigotimes_{j=1}^n \sum_{k=0}^{d-1} \beta_{k}^{(i)} Z^k_j,
\end{equation}
where $\beta_{k}^{(i)}$ are the complex coefficients from expressing $D_j^{(i)}$  in terms of powers of $Z_j$. Rewriting the above expression gives:
\begin{equation}
    \label{eq:h_diag}
    H_\text{diag} = \sum_{\mathbf{k} \in \{0, \dots , d-1\}^n} \left(\sum_{i=1}^{N_\text{params}} \alpha_i \prod_{j=1}^n \beta_{k_j}^{(i)}\right) Z_\mathbf{k},
\end{equation}
which is of the form Eq.~\eqref{eq:u_diag}, where the sum runs over $Z_\mathbf{k}$ with non-zero coefficients. For Eq.~\eqref{eq:z_exp} to be efficient, the number of non-zero terms in Eq.~\eqref{eq:h_diag} must be polynomial in the number of pixels.

\subsection{Model summary}
\label{sec:model_summary}
The model works in two-stages (as with Ref.~\cite{recio2025train}): a train-on-classical stage and deployment on a quantum device. In the following both stages are summarized.

\subsubsection{Train on classical}

To evaluate the loss-function:
\begin{enumerate}
    \item Sample $N_\text{ops}$ operators from Eq.~\eqref{eq:w_m}. This value affects the precision of the MMD estimate.
    \item Sample $N_\text{s}$ strings from a uniform distribution. Each string is of length $n$ and each element takes integer values between $0$ and $d-1$.
    \item Using Eq.~\eqref{eq:z_exp} approximate the expectation values of the $N_\text{ops}$ operators using the $N_\text{s}$ random strings. Note that the statistical error scales as $\sim 1/\sqrt{N_\text{s}}$.
    \item While fixing the samples, the gradient of the loss-function can be found with automatic differentiation~\cite{griewank2008evaluating}.
\end{enumerate}

The loss-function and the gradient can then be used to update the classical optimizer that tunes the parameterized quantum gates in the IQP circuit.

\subsubsection{Deploy on quantum}

Once the model has been classically trained, each image can be generated from sampling the IQP circuit in the computational basis.

In the next section, the IQP qudit model is applied to calorimeter data, which is mapped onto integers with finite precision. In App.~\ref{app:potts}, a second application is analyzed, namely the Potts model~\cite{Wu_82}, at finite temperature. This application illustrates how symmetry constraints can be incorporated into the learning framework.

\section{Application to calorimeter data}
\label{sec:results}

In this section, the main results of our study are presented, focusing on the calorimeter particle energy distribution of Ref.~\cite{belayneh2020calorimetry}. In particular, we investigate how the here proposed method performs in modeling the energy deposition patterns produced by particle showers.

Calorimeter simulations, typically based on Monte Carlo methods like GEANT4~\cite{agostinelli2003geant4}, are computationally expensive. Their outputs are high-dimensional and highly structured and they encode energy deposition patterns that describe the physical processes (electromagnetic shower). Learning this distribution is nontrivial, since it requires capturing local correlations.
This paper uses a publicly available electromagnetic calorimeter (ECAL) electron energy dataset with fixed angle $\theta=\pi/2$ (pseudorapidity 
$\eta =-\log[\tan(\frac{\theta}{2})]=0$) from Ref.~\cite{pierini2020clic}. 
After the dataset is preprocessed (outlined in Sec.~\ref{sec:data_reduction}), the IQP circuit is trained by applying the method described in Sec.~\ref{sec:framework}. The results for the training can be found in Sec.~\ref{sec:iqp_training}, while the validation is deferred to Sec.~\ref{sec:circ_validation}. For a small system, up to $\sim 25$ qubits, the trained parameterized circuit  can be easily sampled. The circuit is implemented in \texttt{PennyLane}. For larger systems, where sampling is not possible, results of the MMD loss function and the covariance matrix with the trained circuits are shown. 
For all experiments, the kernel bandwidth, an additional \textit{scale} factor for the initialization of the parameterized gates, and the optimization learning rate were tuned through a systematic hyperparameter search, and the best-performing configuration was used for the reported results.

\subsection{Data reduction}\label{sec:data_reduction}

\begin{figure}[htp!]
    \centering
    \includegraphics[width=0.97\columnwidth]{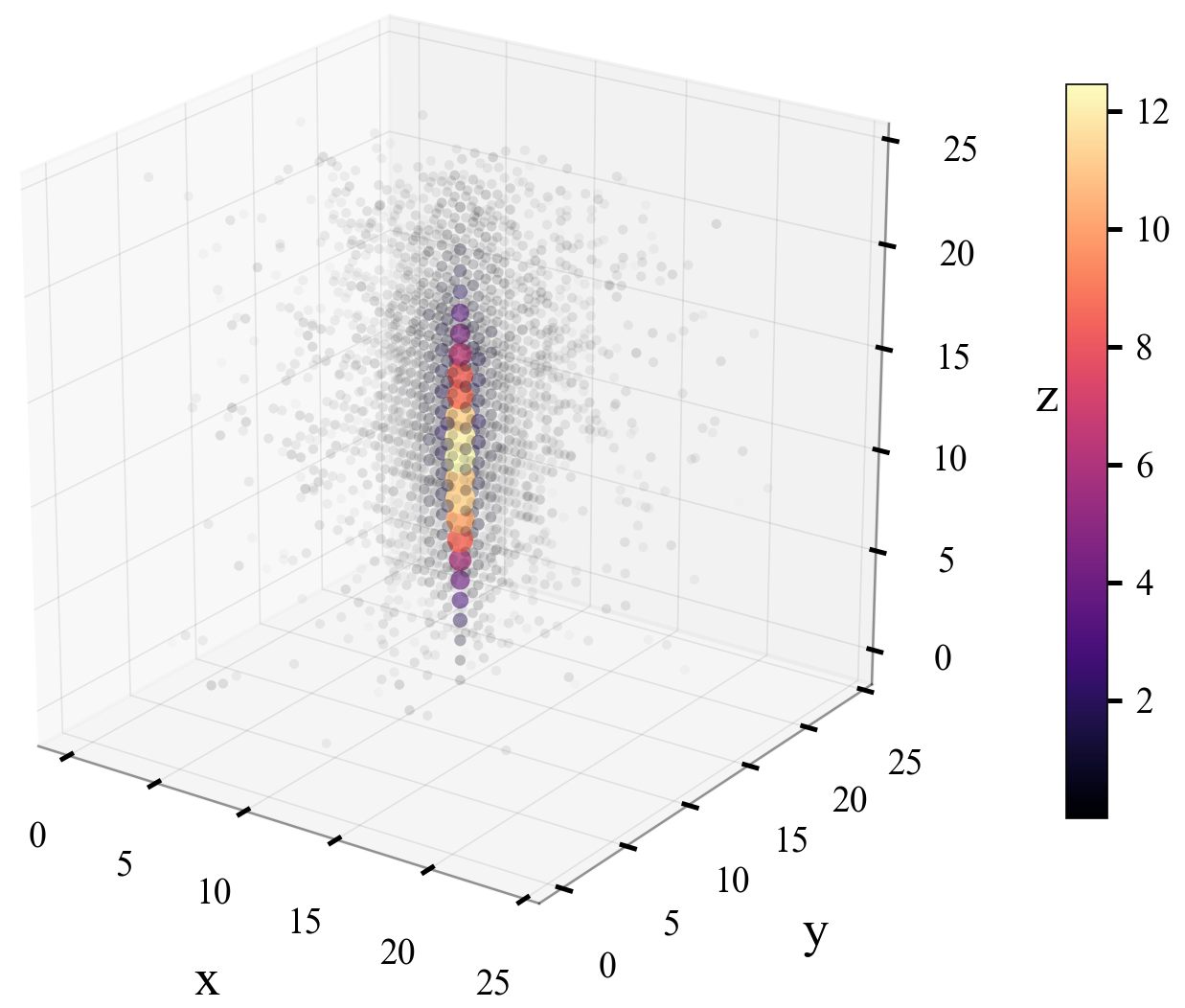}
    \caption{\textbf{3D calorimeter image example:} From the original dataset of $51\times51\times25$ images, a centered window of size $25\times25\times25$ is extracted around the barycenter, while events whose window extends beyond detector boundaries are discarded. (Note that the values correspond to the ECAL energy deposition and not the true incident particles energy.)}
    \label{fig:3d_image_25x25x25_example}
\end{figure}

As a preprocessing step, only events within the energy interval $[225,275]$ GeV are retained for further analysis~\cite{rehm2024precise}. 
Subsequently, each three-dimensional shower image is projected onto the transverse plane by summing over the $z$ direction, from which the energy-weighted barycenter $(x_b,y_b)$ is computed. A centered window of size $25\times25\times25$ is then extracted around this barycenter, while events whose window extends beyond detector boundaries are discarded. 
Fig.~\ref{fig:3d_image_25x25x25_example} depicts an example of a 3D image of the amount of energy deposited along different axes ($x$ and $y$ correspond to the transverse planes and $z$ is the calorimeter depth).

\begin{figure}[htp!]
    \centering
        \begin{subfigure}[t]{0.5\textwidth}
        \centering
        \includegraphics[width=\linewidth]{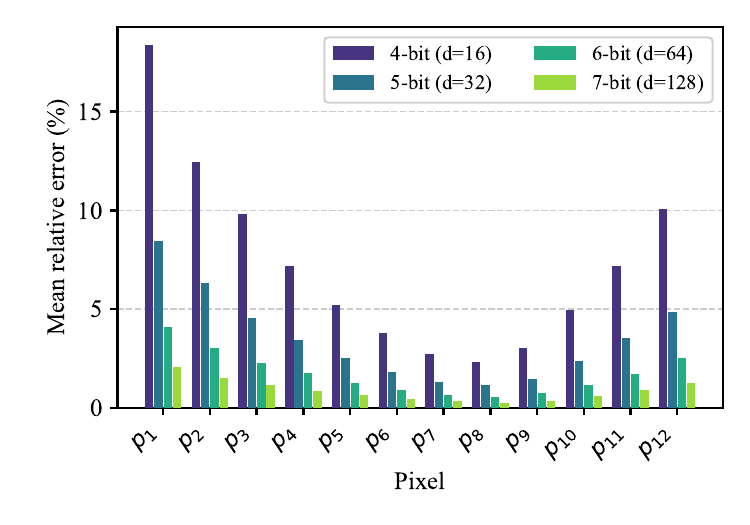}
        \caption{}
        \label{fig:mean_relative_error_f12}
    \end{subfigure}
    \begin{subfigure}[t]{0.48\textwidth}
        \centering
        \includegraphics[width=\linewidth]{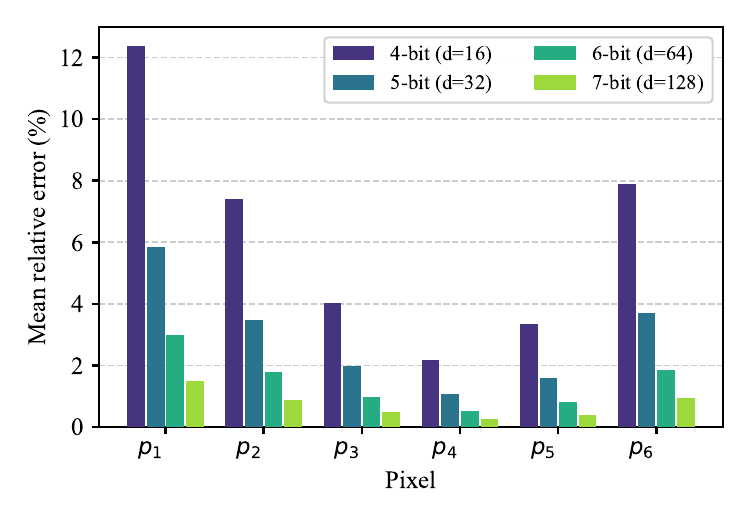}
        \caption{}
        \label{fig:mean_relative_error_f6}
    \end{subfigure}
    \caption{\textbf{Mean relative error between transformed and original data for two pixels reduction:} using Eqs.~\eqref{eq:feature_wise_mean}-\eqref{eq:map_integer}, the accuracy of the transformation to integers with respect to the choice of dimension $d$ can be estimated. The relative error is computed for each of the pixel with a range of $d\in [16,32,64,128]$ for 12 and 6 pixels in panel (a) and (b), respectively.}
    \label{fig:mean_relative_error_f6_f12}
\end{figure}

The cropped 3D shower is reduced to a 1D longitudinal profile ($z$-axis projection) by summing over the two transverse dimensions.
This target dataset is then split into a training and validation set, each consisting of $N_{\text{ts}}\sim 10^3$ samples. 
In this paper we consider a reduced distribution of 6 and 12 pixels. The former allows us to sample from a simulator while the latter example, which requires 60 qubits, shows the scaling possibilities of this algorithm.
In these cases, the model demonstrates stable training and satisfactory performance. 
A similar reduction of this dataset to 8 pixels has been explored in Ref.~\cite{rehm2024precise,monaco2025quantum} with a generative method called Quantum Angle Generator (QAG).
In App.~\ref{app:add_results}, additionally preliminary results for the larger 25-pixel configuration are reported. In this regime, a decrease in correlation accuracy is observed. Addressing this scaling limitation and improving performance with a more extensive research is an important direction for future investigation.
All the experiments were conducted on a laptop equipped with an Apple M5 chip and 16 GB of RAM.

Since the target covariance matrix present small variance among the $n$ pixels, the following transformation is applied to the dataset.
First, for each feature component $k = 1, \dots, n$, the mean and standard
deviation are computed over the $N_{\text{ts}}$ samples of the training set,
\begin{align}\label{eq:feature_wise_mean}
    \mu_k = \frac{1}{N_{\text{ts}}} \sum_{i=1}^{N_{\text{ts}}} x_{i,k},
\qquad
\sigma_k = \sqrt{\frac{1}{N_{\text{ts}}} \sum_{i=1}^{N_{\text{ts}}}
\left(x_{i,k} - \mu_k\right)^2},
\end{align}
where $x_{i,k}$ denotes the $k^\text{th}$ component of the $i^{\text{th}}$ sample. The dataset
is then standardized component-wise,
\begin{equation}\label{eq:map_standardize}
    \tilde{x}_{i,k} = \frac{x_{i,k} - \mu_k}{\sigma_k}.
\end{equation}
After defining a discrete domain with $z_{\min} = 0$ and $z_{\max} = d - 1$,
the standardized data is linearly mapped onto an integer lattice,
\begin{equation}\label{eq:map_integer}
    z_{i,k} = \left\lfloor
(z_{\max} - z_{\min}) \cdot
\frac{\tilde{x}_{i,k} - \tilde{x}_{\min}}{\tilde{x}_{\max} - \tilde{x}_{\min}}
+ z_{\min}
\right\rceil,
\end{equation}
where $\tilde{x}_{\min} = \min_{i,k}\, \tilde{x}_{i,k}$ and
$\tilde{x}_{\max} = \max_{i,k}\, \tilde{x}_{i,k}$ are the global minimum and
maximum taken over all samples $i$ and feature components $k$ of the
standardized dataset, and $\lfloor\,\cdot\,\rceil$ denotes rounding to the
nearest integer.

For each value of $d$, the data is linearly rescaled from its original range to the discrete integer range $[0, d-1]$, effectively discretizing the inputs. The inverse mapping reconstructs approximate values from these integers, and the relative error measures the information loss introduced by this discretization step.
Since the arbitrary value of $d$ defines the accuracy of the integer transformation, as shown with the mean relative error for 12 pixels in Fig.~\ref{fig:mean_relative_error_f12}, we tested the circuits for $d=32$ where the relative error is below $10\%$ for each of the pixels. For the small system with 6 pixels, we consider a value of $d$ such that the circuit can be directly simulated, i.e. $d=16$ giving a total of 24 qubits. For this dimension, the pixels have a mean relative error below $15\%$ Fig.~\ref{fig:mean_relative_error_f6}.

\subsection{IQP circuits training}
\label{sec:iqp_training}
In this section, we describe the variational training of the IQP circuits used as the generative
model, showing the results of the optimization procedure. We will validate our model in Sec.~\ref{sec:circ_validation}.
An important aspect to take into account in the training is the initialization of the circuit parameters $\boldsymbol{\theta }$ in the central layer of Fig.~\ref{fig:q_circuits_qudit}. In Ref.~\cite{lerch2026iqp}, the authors demonstrate that a data-dependent $\boldsymbol{\theta }$ initialization converges faster than an agnostic unbiased strategy with an uniform distribution over all possible configurations.
In this work, the empirical moments of the dataset is used to encode structure directly into the parameters. For single-qudit terms, parameters are derived from the diagonal elements of the covariance matrix. Each diagonal entry is first normalized by the maximum absolute covariance value across all qudits, then mapped into a bounded interval via
\begin{equation}
\theta_i = \arcsin\left(\sqrt{\frac{|\text{Cov}_{ii}|}{\max_{k,\ell}|\text{Cov}_{k\ell}|}}\right),
\end{equation}
ensuring that all single-qubit rotation parameters remain within a stable numerical range.
For two-qubit interaction terms, parameters are constructed from the off-diagonal covariance entries. Each covariance element $\text{Cov}_{ij}$ is normalized by the maximum absolute covariance and scaled by a global \textit{scale} control parameter, which preserves the sign of the correlation while controlling overall interaction strength.

During the training, the convergence is monitored by evaluating the MMD on a fixed validation dataset.
All the results presented here are carried out with the \textit{Adam} classical optimizer~\cite{kingma_2017} at a fixed learning rate of $0.01$. The number of operators sampled to evaluate the MMD was set to $N_\text{ops} = 1000$ and the number of uniformly sampled strings samples used to evaluate the expectation values was set to $N_\text{s} = 1000$. 

\subsubsection{Case study: 6 pixels}
\label{sec:6_train}

\begin{figure}[htp!]
    \centering
    \includegraphics[width=0.94\columnwidth]{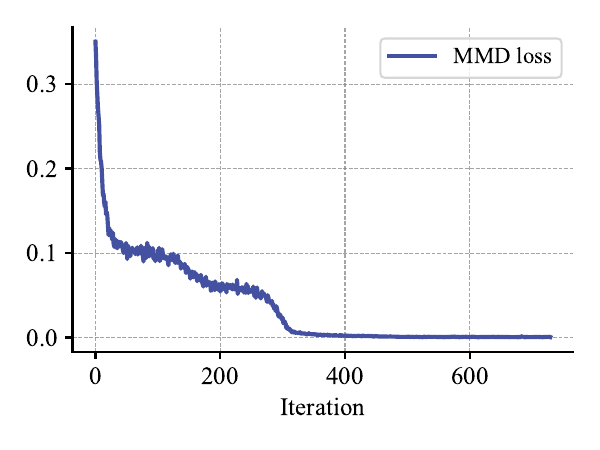}
    \caption{\textbf{Loss function for 6 pixels images and qudit dimension $\mathbf{d=16}$:} The MMD loss function in the IQP circuits training with kernel bandwidth $\sigma=5$ and scale for the initial gates parameters of $0.03$. }
    \label{fig:mmd_f6_d16}
\end{figure}

The first analysis performed was a training on a 6-pixels dataset with qudit dimension $d=16$. The total number of qubits required for this study is 24. This relatively simple case allows us to sample from the trained circuit using a simulator and compare the so-obtained distribution with the target dataset. For larger systems, this procedure is possible only with a real quantum device. A total number of 300 parameters are used in the training. 

Fig.~\ref{fig:mmd_f6_d16} illustrates the evolution of the MMD loss function with a bandwidth of $\sigma = 5$ and an initial gate parameter scale of $0.03$. These hyperparameter values were selected due to their observed performance. To monitor convergence, the MMD was evaluated every 100 iterations on a validation dataset. As can be seen in the figure, the training is relatively smooth. The validation of the model is postponed to Sec.~\ref{sec:6_val}.

\subsubsection{Case study: 12 pixels}
\label{sec:12_train}
\begin{figure}[htp!]
    \centering
    \includegraphics[width=0.94\columnwidth]{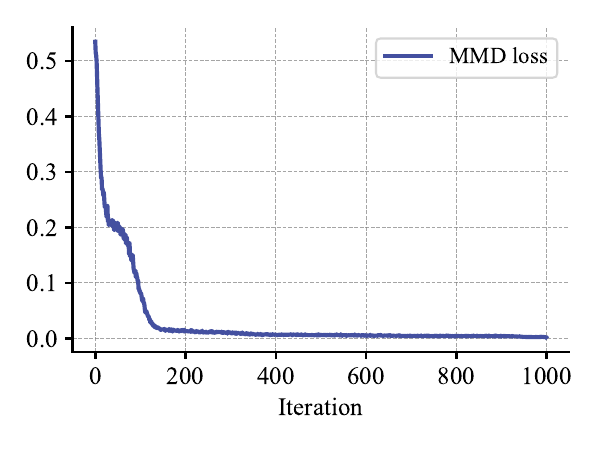}
    \caption{\textbf{Loss function for 12 pixels images and qudit dimension $\mathbf{d=32}$:} The MMD loss function in the IQP circuits training with kernel bandwidth $\sigma=10$ and scale for the initial gates parameters of $0.05$.}
    \label{fig:mmd_f12_d32}
\end{figure}

As a second example of training with the IQP circuit, a larger dataset of 12 pixels with $d=32$ is explored, thus requiring a total of 60 qubits. A case, where quantum hardware is required for sampling the final probability distribution. 
A total number of 1830 variational parameters are used in the training.
As in the previous case, a hyperparameter search for the choice of bandwidth and scale of initial parameters is performed. The results of the best performing configuration are shown here, with $\sigma=10$ and scale $0.05$. Fig.~\ref{fig:mmd_f12_d32} depicts the MMD loss function in the training of the IQP circuit. 
Also in this case, the MMD was evaluated on a validation dataset every 100 iterations, to monitor convergence.

\subsection{IQP circuits validation}
\label{sec:circ_validation}

Having explored the training of the IQP circuit model in the previous section, this section undertakes the validation of these models. This is achieved by examining the discrepancy between the generated distribution and the validation data for different metrics. An example being the covariance. The covariance is used to evaluate how well a model captures the joint variability and dependencies among pixels, providing a statistical check on whether generated samples reflect the structure of the target data.
Since it is possible to estimate expectation values of parameterized IQP
circuits, the covariance between the $i^\text{th}$ and $j^\text{th}$ pixel can be estimated via
\begin{equation}
    \text{Cov}(\hat n_i \hat n_j) = \langle \hat n_i \hat n_j \rangle - \langle \hat n_i \rangle \langle  \hat n_j\rangle,
\end{equation}
where $\hat n$ is the number operator defined for each qudit as,
 \begin{equation}\label{eq:number_op}
     \hat{n} = \sum_{j=0}^{d-1} j \ket{j}\bra{j}.
 \end{equation}
While this method is useful when direct sampling from the quantum circuit is not possible without a quantum hardware, for the smaller case with 6 pixels we directly sample from the generated distribution to compute the covariance matrix to validate the model performance.

 Beyond the covariance, the average pixel value and the $\text{MMD}^2$ are investigated. Where sampling is possible, further metrics such as the Kullback–Leibler (KL)~\cite{kullback1951information} divergence are reported.

\subsubsection{Case study: 6 pixels}
\label{sec:6_val}

\begin{figure}[htp!]
    \centering
    \includegraphics[width=1\columnwidth]{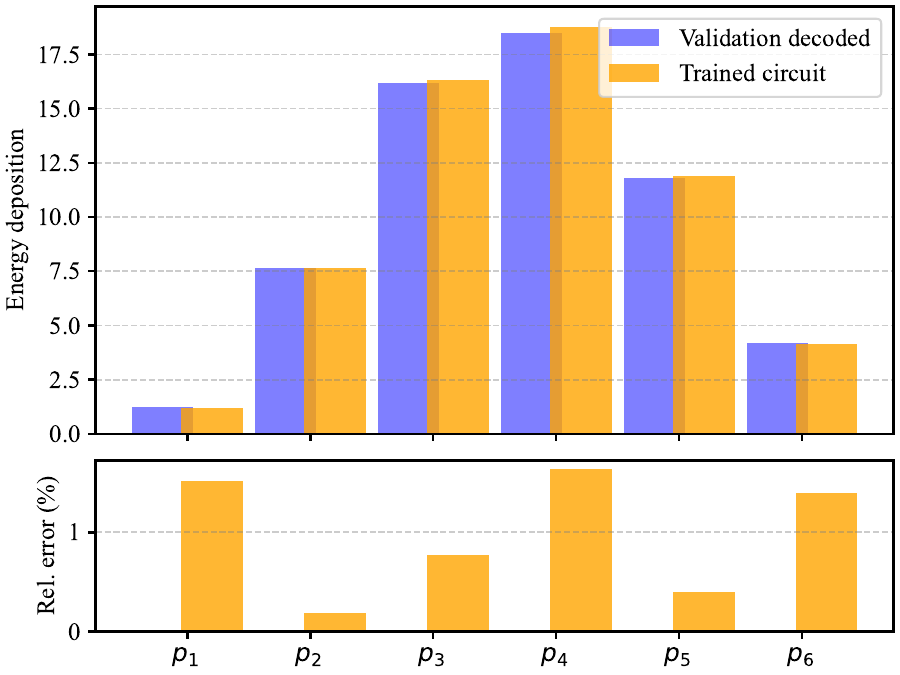}
    \caption{\textbf{Mean energy deposition for 6 pixels images with $\mathbf{d=16}$:} (\textit{top panel}) Energy deposition values of rescaled integer validation distribution (\textit{blue}) and generated samples from the trained IQP circuit (\textit{orange}) with 8192 shots on a \texttt{PennyLane} simulator. (\textit{bottom panel}) The relative error of the means, expressed as a percentage ($\%$), is found to be below $2\%$ for all 6 pixels.}
    \label{fig:samples_f6d16}
\end{figure}

\begin{figure}[htp!]
    \centering
    \includegraphics[width=1\columnwidth]{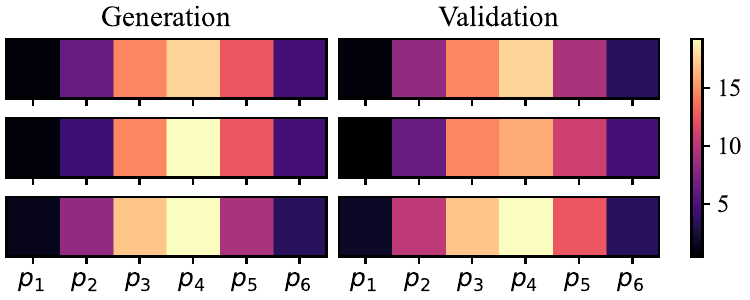}
    \caption{\textbf{Examples of 6 pixels images with $\mathbf{d=16}$:} From the quantum circuit samples with 8192 shots on a simulator, three resulting images are extracted \textit{left panels}) and they are then compared with three images of the validation dataset (\textit{right panels}).}
    \label{fig:samples_f6d16_images}
\end{figure}

\begin{figure*}[ht!]
    \centering
    \begin{subfigure}[t]{0.46\textwidth}
        \centering
        \includegraphics[width=\linewidth]{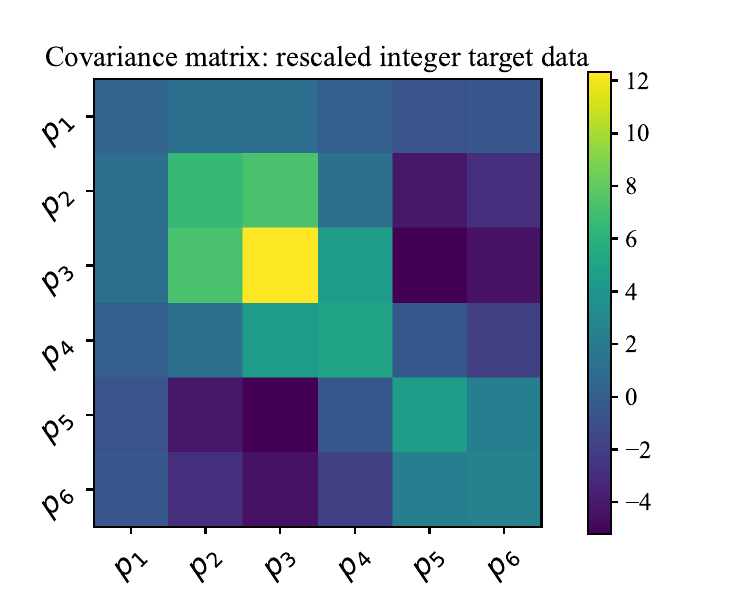}
        \caption{Covariance validation data}
        \label{fig:cov_f6_d16_target}
    \end{subfigure}
    \hfill
    \begin{subfigure}[t]{0.46\textwidth}
        \centering
        \includegraphics[width=\linewidth]{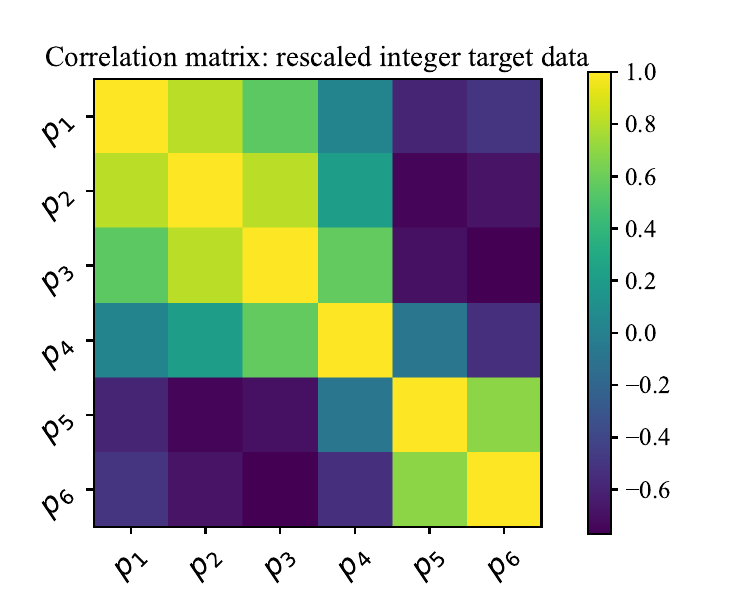}
        \caption{Correlation validation data}
        \label{fig:corr_f6_d16_target}
    \end{subfigure}
    \hfill
    \begin{subfigure}[t]{0.46\textwidth}
        \centering
        \includegraphics[width=\linewidth]{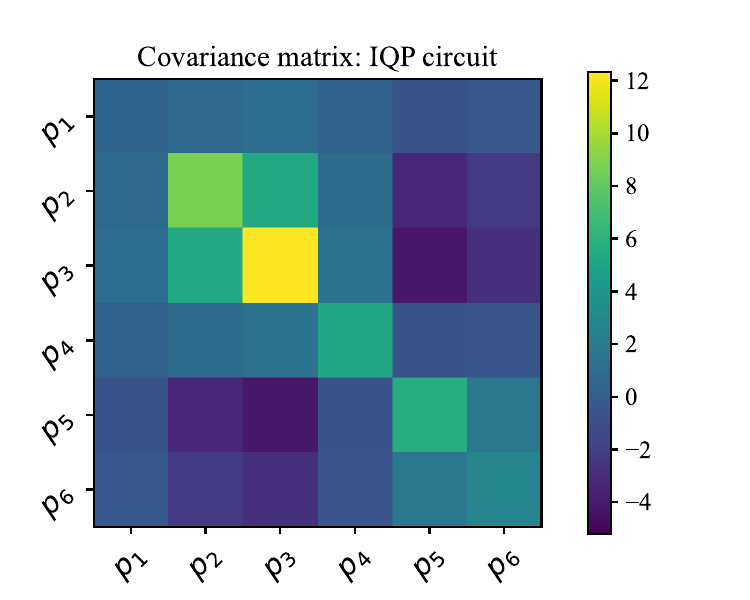}
        \caption{Covariance generation data}
        \label{fig:cov_f6_d16_iqp}
    \end{subfigure}
    \hfill
    \begin{subfigure}[t]{0.46\textwidth}
        \centering
        \includegraphics[width=\linewidth]{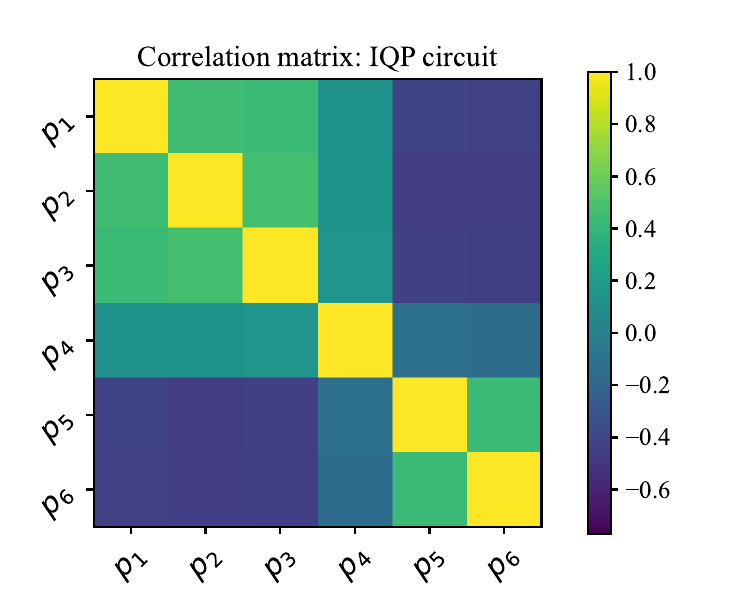}
        \caption{Correlation generation data}
        \label{fig:corr_f6_d16_iqp}
    \end{subfigure}
    \caption{\textbf{Covariance and correlation matrix for 6 pixels images and qudit dimension $\mathbf{d=16}$:} (a) Covariance matrix between pixels of the validation dataset. (b) Correlation matrix between pixels of the same set. (c) Estimated covariance matrix from the IQP circuit sampling with trained parameters, obtained with 8192 samples. (d) Estimated correlation from the IQP circuit, obtained from the covariance calculation. The color scales of the bottom panels are the same as in the top panels. }
    \label{fig:cov_corr_f6_d16}
\end{figure*}

After the training, the model is compared to the withheld validation data, giving a $\text{MMD}^2$ of $ (8.5 \pm 0.7) \times 10^{-4} $. 
The optimized parameters are then used for sampling. The resulting distribution is compared with the validation data. Both are mapped
from the integer values back to the original feature space by inverting the
transformation of Eq.~\eqref{eq:map_integer}. This step is necessary for a fair comparison, as the qudit dimension $d$
controls the discretization precision, as depicted in Fig.~\ref{fig:mean_relative_error_f6_f12}.
The top panel of Fig.~\ref{fig:samples_f6d16} displays, from left to right, the mean energy of the validation distribution (\textit{blue}) and the trained circuit (\textit{orange}). The bottom panel shows the relative error between the mean values, obtained from the trained circuit, and the validation data. The samples were generated using a simulator with 8192 shots. A parameter initialization with $\boldsymbol{\theta }\sim\pi/4$ for single qubit gates was also tested, as in Ref.~\cite{lerch2026iqp}, the final result was however less accurate and thus is not included in this manuscript.

Fig.~\ref{fig:samples_f6d16_images} depicts
three examples of images created via sampling from the IQP quantum circuit with fixed trained parameters. These results (\textit{left panels}) can be compared with the one-dimensional images of the validation dataset. The images generated from the former closely resembles the ones obtained from the latter dataset, indicating good agreement between the two.
Sampling from the quantum circuit allows us to compute the KL divergence between the generation and validation dataset. For the run considered in this manuscript the KL divergence was found to be 0.034, indicating a relatively small distributional shift and good agreement between the two distributions.

Figs.~\ref{fig:cov_f6_d16_target} and~\ref{fig:corr_f6_d16_target} depict the covariance and correlation matrices of the target validation dataset. This dataset is distinct from the training set used for optimization and served as an independent benchmark for evaluating the model performance, including the validation MMD.
Figs.~\ref{fig:cov_f6_d16_iqp},~\ref{fig:corr_f6_d16_iqp} show the covariance and correlation matrices. The calculations use the sampled distribution (Sec.~\ref{sec:6_train}), obtained with the trained quantum circuit. These matrices capture the general structure of the target distribution. 

To quantify the agreement between the validation correlation matrix
($\rho^{\mathrm{val}}$) and the generated correlation matrix
($\rho^{\mathrm{gen}}$), we compute the correlation coefficient
between their entries,
\begin{equation}
r
=
\frac{
\sum_{i<j}^n \left(\rho_{ij}^{\mathrm{val}} - \bar{\rho}^{\mathrm{val}}\right)
\left(\rho_{ij}^{\mathrm{gen}} - \bar{\rho}^{\mathrm{gen}}\right)
}{
\sqrt{
\sum_{i<j}^n \left(\rho_{ij}^{\mathrm{val}} - \bar{\rho}^{\mathrm{val}}\right)^2
\;
\sum_{i<j}^n \left(\rho_{ij}^{\mathrm{gen}} - \bar{\rho}^{\mathrm{gen}}\right)^2
}
},
\end{equation}
where $\rho_{ij}^{\mathrm{val}}$ and $\rho_{ij}^{\mathrm{gen}}$ are the
$(i,j)$ entries of the validation and generated correlation matrices.
The means of the off-diagonal terms in the correlation matrices are denoted by $\bar{\rho}^{\mathrm{val}}$ and $\bar{\rho}^{\mathrm{gen}}$ respectively. For this case we found $r = 0.966$, meaning the model correctly identifies which feature pairs are strongly or weakly correlated. However, the residual element-wise error is on average $|\Lambda| = 0.236$, showing that the IQP circuits produced weaker correlations than the true distribution. A limitation also seen in the binary case in Ref.~\cite{recio2025train}.

\subsubsection{Case study: 12 pixels}
\label{sec:12_val}

\begin{figure}[htp!]
    \centering
    \includegraphics[width=1\columnwidth]{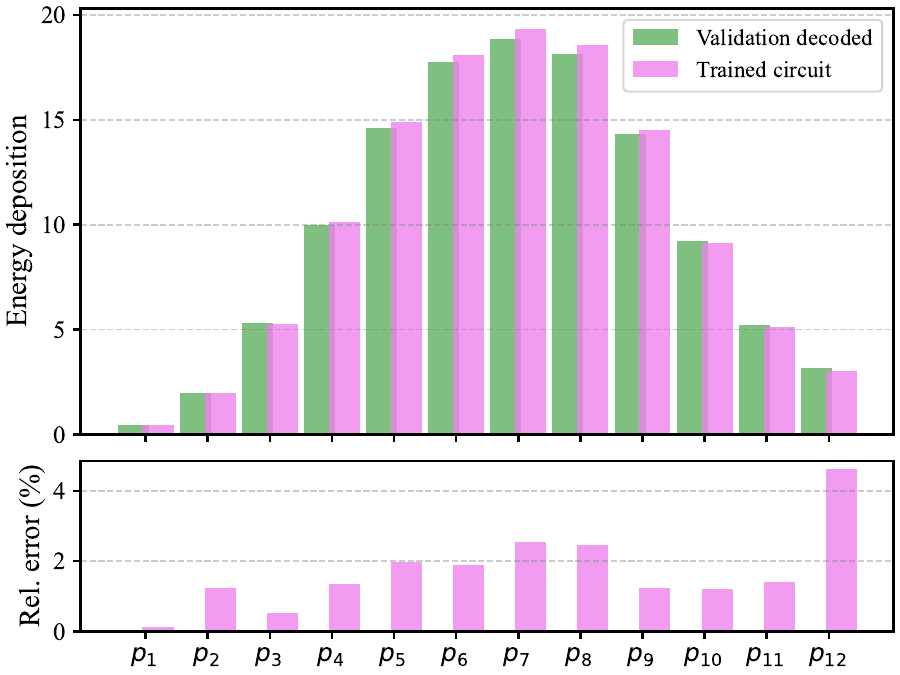}
    \caption{\textbf{Mean energy deposition for 12 pixels images with $\mathbf{d=32}$:} (\textit{top panel}) Energy deposition values of rescaled integer validation distribution (\textit{green}) and the expectation value of the number operator with trained IQP circuit (\textit{pink}) and 1000 samples. (\textit{bottom panel})The relative error of the means, expressed as a percentage ($\%$), depends on the number of samples considered.}
    \label{fig:samples_f12d32}
\end{figure}

The same analysis, as in the previous section, is now repeated for the 12 pixels case. 
The trained model is first compared to the withheld validation dataset. A $\text{MMD}^2$ value of $ (2.9 \pm 0.1)\times 10^{-3} $ is found. 
Since, single images using a simulator cannot be generated in this case , we estimate the expectation value of the number operator for each pixel, Eq.~\eqref{eq:number_op}, and compare the results with the validation dataset in Fig.~\ref{fig:samples_f12d32}. One can see that the average energy deposition obtained with the trained IQP circuit (\textit{pink}) can reproduce the general behavior of the validation data (\textit{green}), with a relative error around $2\%$. The relative error of pixel 12 may reflect finite-sample effects and random fluctuations associated with the particular experimental run.

\begin{figure*}[ht!]
    \centering
    \begin{subfigure}[t]{0.46\textwidth}
        \centering
        \includegraphics[width=\linewidth]{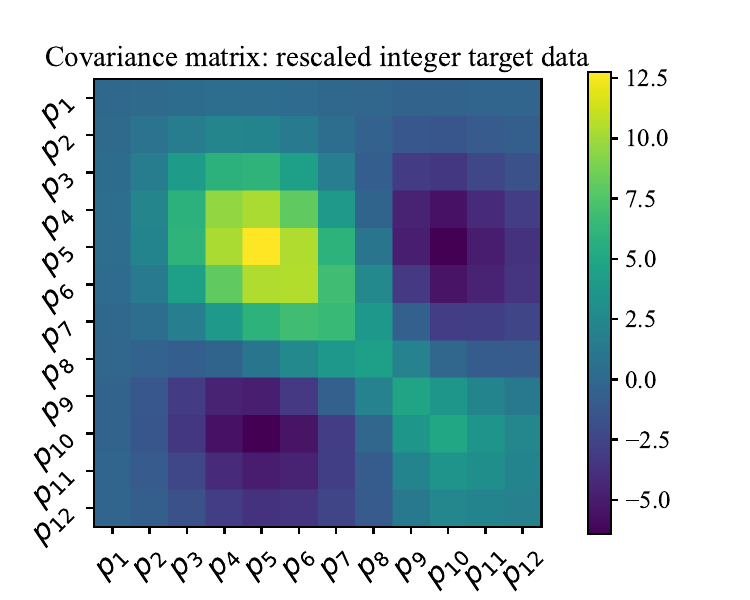}
        \caption{Covariance validation data}
        \label{fig:cov_f12_d32_target}
    \end{subfigure}
    \hfill
    \begin{subfigure}[t]{0.46\textwidth}
        \centering
        \includegraphics[width=\linewidth]{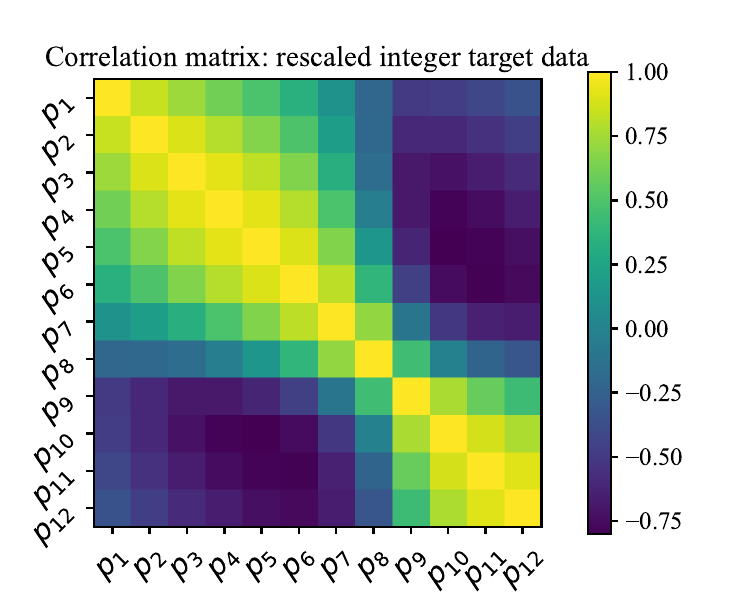}
        \caption{Correlation validation data}
        \label{fig:corr_f12_d32_target}
    \end{subfigure}
    \hfill
    \begin{subfigure}[t]{0.46\textwidth}
        \centering
        \includegraphics[width=\linewidth]{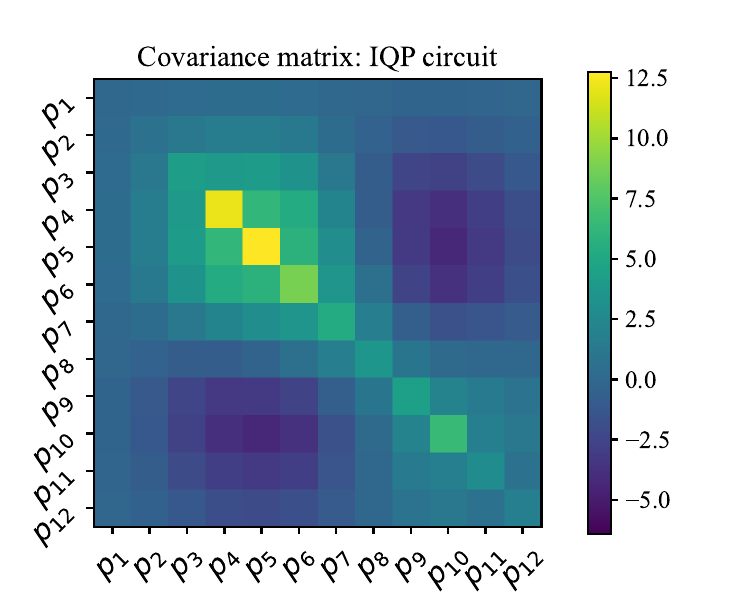}
        \caption{Covariance generation data}
        \label{fig:cov_f12_d32_iqp}
    \end{subfigure}
    \hfill
    \begin{subfigure}[t]{0.46\textwidth}
        \centering
        \includegraphics[width=\linewidth]{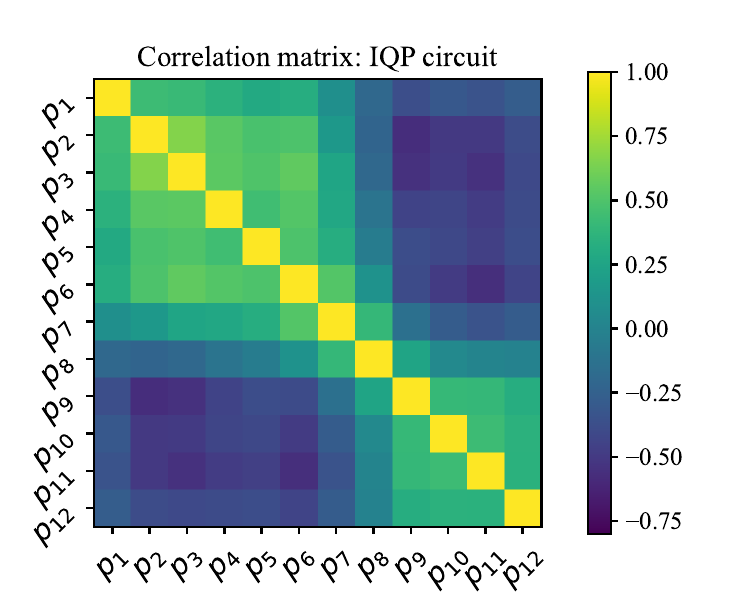}
        \caption{Correlation generation data}
        \label{fig:corr_f12_d32_iqp}
    \end{subfigure}
    \caption{\textbf{Covariance and correlation matrix for 12 pixels images and qudit dimension $\mathbf{d=32}$:} (a) Covariance matrix between pixels of the validation dataset. (b) Correlation matrix between pixels of the same set. (c) Estimated covariance matrix from the IQP circuit, obtained with 1000 samples. (d) Estimated correlation from the IQP circuit, obtained from the covariance calculation. The color scales of the bottom panels  are the same as in the top panels. }
    \label{fig:cov_corr_f12_d32}
\end{figure*}

Fig.~\ref{fig:cov_corr_f12_d32} depicts the covariance and correlation matrices.
In particular, Figs.~\ref{fig:cov_f12_d32_target},~\ref{fig:corr_f12_d32_target} are computed with the validation dataset, while Figs.~\ref{fig:cov_f12_d32_iqp},~\ref{fig:corr_f12_d32_iqp} are obtained from the IQP circuits with the trained parameters.
In the correlation results, the element-wise residual error is on average $|\Lambda| = 0.216$, while the off-diagonal correlation coefficient is $r = 0.974$.

\vspace{0.2cm}

In the experiments presented above, the IQP circuit models captured the overall structure of the target distributions without exhibiting significant mode imbalance. However, the generated distributions showed weaker correlations than those observed in the validation data, as evidenced by the covariance and correlation matrices. This limitation, which was also reported in Ref.~\cite{recio2025train}, suggests that the expressivity of the current ansatz may be too limited to fully reproduce the underlying structure. Future work could address this issue, e.g., by exploring more expressive circuit architectures, alternative gate parameterizations, or loss functions including the correlation.

\section{Circuit implementation}
\label{sec:circ_implement}

The generative model presented in this work can, in principle, be executed on quantum hardware. A central consideration for practical implementation is the efficient transpilation of the circuit. An important aspect that needs to be taken into account when running on a quantum device is qubit connectivity, as certain quantum hardware requires routing to realize interactions between distant qubits. In Ref.~\cite{placidi2026impactqubitconnectivityquantum}, the authors carried out a connectivity-aware analysis of compiled IQP circuits and found that the routing needed on sparse hardware increases circuit depth, pushing implementations closer to the classical simulability threshold. Therefore requiring significantly lower noise rates to maintain an advantage.

The model considered in this paper consists of a QFT acting separately on each qudit at the start of the circuit and correspondingly the inverse QFT ($\text{QFT}^\dagger$) at the end of the circuit. The initial QFT layer can be replaced by a layer of Hadamard gates, since the QFT acts on the all-zero state. The $\text{QFT}^\dagger$ can be realized using a semi-classical scheme when mid-circuit measurements are available~\cite{Griffiths_1996}. In this setting, measurements are deferred and incorporated sequentially, allowing qubits to be measured one at a time and their outcomes classically fed forward to control subsequent operations. This approach reduces circuit depth and eliminates the need for two-qubit gates. 
When fully-coherent evolution is required and mid-circuit measurements with feed-forward are not available, a unitary implementation of the $\text{QFT}^\dagger$ becomes necessary. In these cases, compilation strategies such as Parity Twine \cite{dreier2025connectivityawaresynthesisquantumalgorithms} can be used to reduce connectivity and gate overhead. With the Parity Twine approach, the resulting QFT depth scales as $\nu \log_2 d $, where $d$ is the dimension of the qudits considered and $\nu$ is a constant that reflects the device connectivity (for all-to-all device connectivity $\nu = 2$). Parity Twine can also be used to provide a more hardware-efficient decomposition of the diagonal ansatz in IQP circuits. The circuit-depth of the central parameterized layer, consisting of all one- and two-body qubit $Z$ rotations, scales as $\nu n \log_2 d $, with $n$ the total number of pixels. Therefore, the circuit depth scales linearly with the number of features and logarithmically in the size of each feature. 

It has been shown in Ref.~\cite{cao2026measurement}, that the introduction of mid-circuit measurements in constant-depth IQP circuits allows for richer entanglement structures. Hence, the introduction of mid-circuit measurements in the diagonal ansatz might also provide a way of reducing circuit depth, while maintaining performance.

\section{Discussion}
\label{sec:discussion}

Instantaneous Quantum Polynomial circuits have attracted considerable attention in recent years due to their computational properties and their role in generative quantum machine learning.
They have found application in many research fields, such as mRNA folding~\cite{kumar2025towards} or learning graph distributions~\cite{ballo2025generative}.
IQP circuits have also been considered in a shallow formulation in Refs.~\cite{ballo2026shallow,slim2026iqpbornmachinecalorimeter}. The challenges associated with shallow IQP circuits are discussed in App.~\ref{app:iqp_sparse}.

In this work, a possible extension of IQP circuits for generating non-binary distributions has been presented. By extending the circuit design to a qudit-based formalism we were able to develop a setup that can be used in generative machine learning applications on non-binary datasets. 
The quantum circuit has been adapted to encode each integer valued pixel into a bit-string of fixed length $b$. Therefore, the circuit can be considered as constituted of qudits with dimension $2^b$. The quantum gates have been transformed to the qudit formalism, by replacing the layer of Hadamards with a QFT, acting separately on each qudit. The central diagonal layer follows a similar procedure and a method to estimate the expectation values of Pauli-$Z$ operators has been carried out. 
To develop a generative machine learning model in this framework, the MMD loss function has been modified to be consistent with the proposed formulation.

To validate our method, the model was trained on the energy deposits from single-particle electron showers recorded by the calorimeter in the CLIC particle detector. Accurately modeling these complex, correlated distributions is computationally expensive using classical Monte Carlo methods and quantum computing may be an alternative.
The qudit parameterized IQP model was trained on two reduced datasets of the $z$-axis projection of the electromagnetic calorimeter energy with 6 and 12 pixels, respectively. The first case required only 24 qubits to be implemented allowing for sampling to be simulated, and therefore demonstrate the accuracy of the results. The second case, requiring 60 qubits, belongs to the regime where direct numerical simulation is prohibitive, necessitating the use of a quantum device to sample from the generated distribution.
In the latter case, it was possible to estimate the covariance and correlation matrix, based only on Monte Carlo sampling, testing the method validity.
The IQP circuit model was able to reproduce the overall structure of the target distributions. On the downside, it only partially captured the correlations observed in the validation data. Improving performance may require more expressive circuit designs, alternative gate parameterizations, or correlation-aware loss functions.

Recent work on IQP-based Quantum Circuit Born Machines (QCBMs) has shown that universality can be achieved via auxiliary qubits~\cite{kurkin2025note}, with extensions to linear-optical Boson Sampling Born Machines~\cite{kurkin2026universalityclassicallytrainablequantumdeployed}. 
In Ref.~\cite{kurkin2025universality}, the authors proved that adding $n+1$ auxiliary qubits makes the generative model universal, and by introducing a kernel-adaptive MMD training scheme they improved performance, overcoming limitations of fixed-kernel training. These results suggest a possible direction for improving the present approach, by incorporating similar strategies to enhance the model expressivity.
Beyond applications and expressivity studies, recent work has focused on the limitations of quantum generative models, including trainability barriers associated with anticoncentration~\cite{herbst2025limits}, barren plateaus~\cite{shen2026characterizingtrainabilityinstantaneousquantum, deluca2026trainabilityiqpquantumcircuit} and discrepancies between classical surrogate training and quantum deployment~\cite{herrero2025born}.

Sec.~\ref{sec:circ_implement} discussed possible strategies to implement IQP circuits on quantum hardware, either using the Parity Twine method or mid-circuit measurements. In the presence of amplitude-damping noise, Ref.~\cite{shravan2026efficientsimulationnoisyiqp} presents an efficient classical algorithm for sampling from IQP circuits by exploiting the tendency of the noise to drive the quantum state toward a fixed point, thus enabling an accurate approximation. This demonstrates the need to minimize the noise present in IQP circuits. To this end, error correction techniques must be developed, e.g. Ref.~\cite{hangleiter2025fault} proposes a hardware-efficient framework for implementing classically hard binary IQP sampling circuits on neutral-atom quantum processors using co-designed error correcting codes. Any physical implementation of the integer IQP circuits must also take noise into account, thus motivating further work in this direction.
Binary IQP circuits have also been adapted to other hardware platforms. Implementations of photonic quantum generative model have been explored in Refs.~\cite{gottlieb2026efficienttrainingphotonicquantum,kolarovszki2026generativemodelinggaussianboson,kurkin2026universalityclassicallytrainablequantumdeployed}. While, in Ref.~\cite{bako2025fermionic}, the authors present a classically trainable quantum generative model based on fermionic linear optical circuits and parameterized magic states. Future work might examine extending these ideas to the integer IQP circuits case, particularly in platforms where two-level systems are not native.

In conclusion, our work illustrates a potential extension of parameterized IQP circuits, highlighting their role as a framework for quantum generative modeling with integer data and evaluating their capabilities and performance.

\section*{Acknowledgments}
This work was developed within the project Hamburg Full Stack Quantum Machine Learning (HQML) in collaboration with consortium partners DESY, eleQtron and the DLR Quantum Computing Initiative (DLR QCI) and funded by the Hamburgische Investitions- und F\"orderbank (IFB). We would like to thank Saverio Monaco for fruitful discussions.

\section*{Data availability}
Data to reproduce the plots presented in this study are publicly available on Zenodo~\cite{banks_2026_20544613}.

\appendix

\section{Pauli qudit operator identities}
\label{app:decomp}
This section reports qudit-based operators definitions and related identities used in the formulation and analysis of the model.

\subsubsection{Useful qudit identities}
Below presents some useful relations between the qudit operators:
\begin{subequations}
\begin{align}
    H_d Z_d H_d^\dagger  & = X_d^{-1},
    \\
    H_d X_d H_d^\dagger  & = Z_d,
     \\
    Z_d^p X_d^q &= \omega^{pq} X_d^qZ_d^p,
     \\
    H_d Z_d^k H_d^\dagger  & = X_d^{d-k},
\end{align}
\end{subequations}
where $\omega=e^{\frac{2 \pi i }{d}}$.

\subsubsection{List of some of the standard operators in terms of Pauli qudit $Z$}

Here, decompositions of the operators used in this paper into Pauli-$Z$ operators for qudits are provided. Note that all operators act on $d$-dimensional qudits.
A projector onto a computational basis state:
\begin{equation}
    \ket{j}\bra{j} = \frac{1}{d}\sum_{k=0}^{d-1} \omega^{-jk} Z^k.
\end{equation}

The number operator:
 \begin{equation}
     \hat{n} = \sum_{j=0}^{d-1} j \ket{j}\bra{j} = \frac{d-1}{2} I -\sum_{m=1}^{d-1} \frac{Z^m}{1-\omega^{-m}}.
 \end{equation}

The number operator squared:
\begin{subequations}
 \begin{align}
     \hat{n}^2 & = \sum_{j=0}^{d-1} j^2 \ket{j}\bra{j}\\ &= \frac{1}{6}(d-1)(2d-1) I \nonumber \\
     & \hspace{1 cm} -\sum_{m=1}^{d-1} \left(\frac{d}{1-\omega^{-m}} + \frac{2 \omega^{-m}}{(1-\omega^{-m})^2}\right)Z^m.
 \end{align}
\end{subequations}
The product between number operators computed onto the $i^{\text{th}}$ and $j^{\text{th}}$ qudit:
\begin{align}
\hat n_i \hat n_j 
&=
\frac{(d-1)^2}{4}I
-
\frac{d-1}{2}
\sum_{m=1}^{d-1}
\left(
\frac{Z_i^{m}}{1-\omega^{-m}}
+
\frac{ Z_j^{m} }{1-\omega^{-m}}
\right)
\nonumber\\
&\quad
+
\sum_{m=1}^{d-1}
\sum_{m'=1}^{d-1}
\frac{ Z_i^{m} Z_j^{m'}}
{(1-\omega^{-m})(1-\omega^{-m'})}.
\end{align}

A projector onto two-states, labelled by $r$ for qudit 1 and $s$ for qudit 2, which satisfy  $k = r-s \bmod d$, is given by:
\begin{equation}
    \label{eq:proj_cyc}
    P_k =\frac{1}{d} \sum_{m=0}^{d-1} \omega^{-km}\left(Z_1Z_2^\dagger\right)^m.
\end{equation}

In the case of $d = 2^b$ (with $b$ bit-string length), the qubit Pauli-$Z$ (denoted by $z_i$) can be expressed in terms of qudit operators. We adopt the convention that qubit 0 is the most significant bit,
\begin{equation}
    z_i = \frac{1}{d}\sum_{r=0}^{d-1} \left[\left(1-\omega^{-r 2^{b-1-i}}\right)\prod_{\substack{j=0\\j \neq i }}^{b-1}\left(1+\omega^{-r 2^{b-1-j}}\right) \right] Z^r,
\end{equation}
or for $z_{\mathbf{g}}$ where $\mathbf{g}$ is a binary vector of length $b$ that denotes where each $z$ acts:
\begin{equation}
    z_{\mathbf{g}} = \frac{1}{d}\sum_{r=0}^{d-1} \left[\prod_{j=0}^{b-1}\left(1+(-1)^{g_j}\omega^{-r 2^{b-1-j}}\right) \right] Z^r.
\end{equation}

\section{Loss-function details}
\label{app:loss_function}

A standard strategy for training generative models is to specify a differentiable objective function that measures model performance on the training data, and then estimate the parameters using gradient-based optimization.
The loss function in this paper is based on the Maximum Mean Discrepancy (MMD)~\cite{gretton2012kernel}, a type of distance function between two distributions. This section elaborates on the challenges in using this approach.
The MMD is based off using a kernel,
\begin{align}
    \hat{k} =\sum_{\boldsymbol{x}, \boldsymbol{y}} k(\boldsymbol{x}, \boldsymbol{y}) \ket{\boldsymbol{x}}\bra{\boldsymbol{x}} \otimes \ket{\boldsymbol{y}}\bra{\boldsymbol{y}},
\end{align}
where $\boldsymbol{x},\boldsymbol{y}$ denotes data stored on two different registers. Setting the kernel to be a Gaussian kernel gives:
\begin{align}\label{eq:kernel}
    \hat{k} & =\sum_{\boldsymbol{x}, \boldsymbol{y}} e^{-\frac{\abs{\boldsymbol{x} -\boldsymbol{y}}^2}{2 \sigma^2}}\ket{\boldsymbol{x}}\bra{\boldsymbol{x}} \otimes \ket{\boldsymbol{y}}\bra{\boldsymbol{y}}\nonumber\\
    & =\sum_{\boldsymbol{x}, \boldsymbol{y}} \prod_{i=1}^n e^{-\frac{\abs{x_i - y_i }^2}{2 \sigma^2}}\ket{x_i}\bra{x_i} \otimes \ket{y_i}\bra{y_i} \nonumber\\
    & = \prod_{i=1}^n \sum_{x_i, y_i} e^{-\frac{\abs{x_i - y_i }^2}{2 \sigma^2}}\ket{x_i}\bra{x_i} \otimes \ket{y_i}\bra{y_i},
\end{align}
with $\sigma$ the kernel bandwidth. This value reflects the distances of the target data, in Ref.~\cite{recio2025train} the median value of pairwise distances of the points is often considered. The term $\abs{\cdot}^2$ denotes length of its input, it is assumed that this can be evaluated on each individual component of a vector input and then summed over.
From Eq.~\eqref{eq:kernel} it is clear that the kernel depends only on the distance between $x_i$ and $y_i$. Therefore it can be rewritten as 
\begin{equation}
    \label{eq:kernel_projector}
    \hat{k} = \prod_{i=1}^n \sum_{k\in \text{distances}} e^{-\frac{k^2}{2 \sigma^2}} P_{\abs{x_i-y_i}=k},
\end{equation}
where $P_{\abs{x_i-y_i}=k}$ is the projector onto states that satisfy $\abs{x_i-y_i}=k$. To make progress, one needs to decide on a distance. A possible choice could be the Euclidean distance, where
\begin{multline}
    P_{\abs{x_i-y_i}=k}^\text{Euc.} \\
    = \frac{1}{d^2}\sum_{p=0}^{d-1}\sum_{q=0}^{d-1}\left[\sum_{r=0}^{d-1}\sum_{s=0}^{d-1} \omega^{-pr}\omega^{-qs} \delta_{\abs{r-s}=k}
    \right] Z^p_{x_i} Z^q_{y_i},
\end{multline}
with $\delta$ the Kronecker-delta. The double sum over $p$ and $q$ complicates the evaluation of the loss-function. Ultimately, we would like to interpret the coefficient in front of each power of $Z$ as a probability to perform Monte Carlo sampling. Having all powers of $Z$ on both qudits requires working over a probability distribution with $d^2$ outcomes. This may become prohibitive at large $d$. Instead we define a \textit{cyclic} distance between qudit states:

\begin{equation}
    P_k^\text{Cyc.} =\frac{1}{d} \sum_{m=0}^{d-1} \omega^{-km}\left(Z_{x_i}Z_{y_i}^\dagger\right)^m,
\end{equation}
which projects onto states that satisfy $x_i-y_i = k \bmod d$. To compensate the one-sidedness of this projector the distance is defined as $\min(k,d-k)$. For example if $x_i = 3$ and $y_i = 10$, then $x_i - y_i = d-7$ and the distance should be $\min(d-7,7)$. Hence the kernel operator becomes:

\begin{align}
    \hat{k} & = \prod_{i=1}^n \sum_{k=0}^{d-1} e^{-\frac{\min(k, d-k)^2}{2 \sigma^2}} P_k^\text{Cyc.} \nonumber\\ 
    & = \prod_{i=1}^n  \frac{1}{d} \sum_{k=0}^{d-1} e^{-\frac{\min^2(k,d-k)}{2 \sigma^2}}\sum_{m=0}^{d-1} \omega^{-km} (Z_{x_i} Z_{y_i}^\dag)^m\nonumber\\
    & = \prod_{i=1}^n  \sum_{m=0}^{d-1} \left(\frac{1}{d} \sum_{k=0}^{d-1} e^{-\frac{\min^2(k,d-k)}{2 \sigma^2}} \omega^{-km} \right) (Z_{x_i} Z_{y_i}^\dag)^m\nonumber\\
    & = \prod_{i=1}^n   \sum_{m=0}^{d-1} w_{\sigma}(m) (Z_{x_i} Z_{y_i}^\dag)^m,
\end{align}
where $w_{\sigma}(m)$ is defined as
\begin{equation}
    w_{\sigma}(m) = \left(\frac{1}{d} \sum_{k=0}^{d-1} e^{-\frac{\min^2(k,d-k)}{2 \sigma^2}} \omega^{-km} \right).
\end{equation}
With $\mathcal{Z} = \sum_m \abs{w_\sigma(m)}$ and $p_\sigma (m) = \abs{w_\sigma(m)} / \mathcal{Z}$, then the kernel can be written as:
\begin{align}
    \hat{k} = & \sum_{m_1,\ldots,m_n = 0}^{d-1}
     \prod_{i=1}^n \mathcal{Z} \frac{p_\sigma(m_i)}{\abs{w_\sigma(m_i)}} w_{\sigma}(m_i)\, (Z_{x_i} Z_{y_i}^\dag)^{(m_i)} \nonumber \\
     = &  \mathcal{Z}^n \sum_{m_1,\ldots,m_n = 0}^{d-1}
     \prod_{i=1}^n   p_\sigma(m_i) \text{sign}(w_{\sigma}(m_i))\, (Z_{x_i} Z_{y_i}^\dag)^{(m_i)} \nonumber \\
      = &  \mathcal{Z}^n \mathbb{E}_{m_i\sim p_\sigma}
      \prod_{i=1}^n   \text{sign}(w_{\sigma}(m_i))\, (Z_{x_i} Z_{y_i}^\dag)^{(m_i)}\label{eq:mmd_iqp}.
\end{align}

It is straightforward to show that $w_{\sigma}(m)$ are real and sum to one. However, they are not guaranteed to be positive, meaning that $\mathcal{Z} \geq 1$, making Monte Carlo methods inefficient. One way around this limitation is to work in the limit $\sigma \ll d$, in this limit $\mathcal{Z} = 1$. Alternatively, in the main body we take simply a probability distribution (Eq.~\eqref{eq:w_m}), inspired by taking the Fourier transform (over the reals) of a continuous Gaussian with bandwidth $\sigma$. This case is positive everywhere. Future work could consider other kernel choices or adding a sufficient number of auxiliary qudits to each qudit to work in the regime $\sigma \ll d$.

\section{Incorporating symmetries into the model}
\label{app:potts}
In Ref.~\cite{recio2025train}, the authors demonstrate how a $\mathbb{Z}_2$ symmetry in the data can be incorporated into a binary IQP ansatz. They utilize the resulting ansatz to train a circuit for sampling an Ising model at finite temperature. In this section, it is shown how to incorporate a permutation symmetry into the qudit based IQP circuit. Then the performance of the proposed method is evaluated by training a circuit to sample from the Potts model at finite temperature. The Potts model is a generalization of the Ising model to higher-dimensional spin variables.

\subsection{Estimating expectation values}

Given data $\boldsymbol{x}$ labelled by integers $0, \dots, d-1$, a permutation $s$ can be applied to the labels of the data. An example permutation could be $0 \rightarrow 1, 1\rightarrow2, \dots, d-1   \rightarrow 0 $. We desire a model that has the property
\begin{equation}
    p_\theta(\boldsymbol{x}) = p_\theta(s(\boldsymbol{x})).
\end{equation}
This can be achieved by modifying the initial state of the IQP qudit-based model to be invariant under the permutation $s$, in particular by choosing a GHZ-like state defined as follows:
\begin{equation}
    \label{eq:ghz}
    \ket{\psi_i} = \frac{1}{\sqrt{d}} \sum_{i=0}^{d-1} \ket{i}^{\otimes{n}}.
\end{equation}
As with the $\mathbb{Z}_2$, changing the initial-state requires modifying how expectation values are calculated. This is carried out for the permutation case below, writing the IQP qudit ansatz as $U_{\text{diag}}$, previously defined in Eq.~\eqref{eq:u_diag},
\begin{widetext}
\begin{subequations}
\begin{align}
    & \bra{\mathbf{p}} Z_\mathbf{a} \ket{\mathbf{q}} =  \bra{\mathbf{p}} H^{\otimes n, \dagger} U_{\text{diag}}^\dagger H^{\otimes n} Z_{\boldsymbol{a}}  H^{\otimes n, \dagger} U_{\text{diag}} H^{\otimes n} \ket{\mathbf{q}} \\ 
    & =\frac{1}{d^n} \sum_{\boldsymbol{z}}  \omega^{\boldsymbol{z}\cdot \left(\mathbf{q}-\mathbf{p}\right)+\mathbf{p}\cdot \mathbf{a}}\exp \left[i \sum_j \left(\theta_j \omega^{\mathbf{g_j}\cdot \boldsymbol{z}} - \theta_j^*\omega^{-\mathbf{g_j} (\boldsymbol{z}-\mathbf{a})}\right) \right].
\end{align}
\end{subequations}
The above expression can be simplified by making use of the fact $\mathbf{p} = p (1,1,\dots 1)^T$ and likewise for $\mathbf{q}$.
Thus,
\begin{subequations}
    \begin{align}
    \langle Z_\mathbf{a} \rangle & =\sum_{\boldsymbol{z}}\frac{1}{d}\sum_{p=0}^{d-1}\sum_{q=0}^{d-1} \frac{\omega^{p \| \mathbf{a} \|_1}}{d^n}   \omega^{ - \left(p-q\right)\| \mathbf{z} \|_1}\exp \left[i \sum_j \left(\theta_j \omega^{\mathbf{g_j}\cdot \mathbf{z}} - \theta_j^* \omega^{-\mathbf{g_j} (\mathbf{z}-\mathbf{a})}\right) \right]\\
    &=\frac{1}{d^n}\sum_{\boldsymbol{z}}\exp \left[i \sum_j \left(\theta_j \omega^{\mathbf{g_j}\cdot \mathbf{z}} - \theta_j^* \omega^{-\mathbf{g_j} (\mathbf{z}-\mathbf{a})}\right) \right]\frac{1}{d}\sum_{p=0}^{d-1} \omega^{p (\| \mathbf{a} \|_1 - \| \mathbf{z} \|_1)} \sum_{q=0}^{d-1}    \omega^{q \| \mathbf{z} \|_1}\\
    \label{eq:exp_perm}
    &= d \, \delta_{\| \mathbf{a} \|_1}\, \mathbb{E}_{\mathbf{z}\sim U} \exp \left[i \sum_j \left(\theta_j \omega^{\mathbf{g_j}\cdot \mathbf{z}} - \theta_j^* \omega^{-\mathbf{g_j} (\mathbf{z}-\mathbf{a})}\right) \right] \delta_{\| \mathbf{z} \|_1} ,
\end{align}
\end{subequations}
\end{widetext}
where $\|\cdot \|_1$ denotes the standard vector 1-norm and $\delta_k$ denotes the Kronecker-delta, which assumes the value 1 if $k \bmod d=0$ and $0$ otherwise. Hence, when incorporating permutation symmetry into the model, Eq.~\eqref{eq:exp_perm} is used to estimate expectation values.

\subsection{A suitable loss-function}

The MMD relies on a kernel function; here, a kernel that better incorporates the categorical nature of the data labels is constructed. Eq.~\eqref{eq:kernel_projector} shows a kernel operator acting between data encoded in two registers. The distance between the states is defined to be zero if they are the same and 1 otherwise, this gives the kernel:
\begin{align}
    \hat{k} & = \prod_{i=1}^f  P_{0}^{x_i,y_i} + e^{-\frac{1}{2\sigma^2}} \left(I^{x_i,y_i}-\prod_{i=1}^f  P_{0}^{x_i,y_i}\right)\nonumber\\
     & = \prod_{i=1}^f e^{-\frac{1}{2\sigma^2}}I^{x_i,y_i}+ \left(1-e^{-\frac{1}{2\sigma^2}}\right)   P_{0}^{x_i,y_i},
\end{align}
where $P_{0}^{x_i,y_i}$ is the projector onto states that satisfy $\abs{x_i-y_i}=0 \mod d$. Substituting in $P_{0}$ in terms of Pauli Z operators (Eq.~\eqref{eq:proj_cyc}) gives:
\begin{align}
    \hat{k} &  = \sum_{m_1,m_2\dots m_n=0}^{d-1}\prod_{i=1}^n  w_\sigma(m_i)\left(Z_{x_i}Z_{y_i}^\dagger\right)^{m_i}\nonumber\\
     & = \mathbb{E}_{m_i\sim w_\sigma}
 \prod_{i=1}^n (Z_{x_i} Z_{y_i}^\dag)^{(m_i)},
\end{align}    
where
\begin{equation}
    w_\sigma(m) =
    \begin{cases}
        e^{-\frac{1}{2\sigma^2}}+ \frac{1}{d}\left(1-e^{-\frac{1}{2\sigma^2}}\right) & \text{ if }m =0 \\
        \frac{1}{d}\left(1-e^{-\frac{1}{2\sigma^2}}\right) & \text{ otherwise.}
    \end{cases}
\end{equation}
The $w_\sigma(m)$ are all positive and sum to 1 and are therefore a valid probability density function. The average weight $\mu$ of an operator in $\hat{k}$ is therefore:
\begin{equation}
    \mu = (1-w_{\sigma}(0))n.
\end{equation}
Consequently, the bandwidth can be expressed in terms of $\mu$,
\begin{equation}
     \label{eq:mu_bandwith}
     \sigma^2 = \frac{-1}{2\log \left( \frac{1-\frac{\mu}{n}-\frac{1}{d}}{1-\frac{1}{d}}\right)}.
\end{equation}

\subsection{The Potts model}
\label{app:potts_res}
To assess the performance of the method, we consider the Potts model~\cite{Wu_82} at finite temperature. The Hamiltonian for the Potts model is given by
\begin{equation}
    \label{eq:potts}
    H_{\text{Potts}} = -\sum_{\langle i, j \rangle} J_{i,j}\mathbbm{1}(S_i=S_j),
\end{equation}
and is defined on a square lattice of size $L\times L$ with periodic boundary conditions. Each $J_{i,j}$ is randomly sampled from a uniform distribution between $0$ and $2$ and each local spin $S$ has dimension $d$. The term $\mathbbm{1}(S_i=S_j)$ is 1 if $S_i=S_j$ and is zero otherwise, hence the Hamiltonian is invariant under a relabelling of the spins. In this paper, we take $L=5$ and $d=16$. Therefore, the IQP circuit involves 25 qudits, encoded into 100 qubits.

\subsubsection{Ansatz}
The diagonal ansatz is taken to be: 
\begin{equation}
    U_\text{diag} = e^{i \sum_{i<j}^n \alpha_{i,j}P_0^{i,j}},
\end{equation}
where $\alpha_{i,j}$ are trainable parameters and $P_0^{i,j}$ is the projector onto computational states with the same value. This choice of ansatz is inspired by the Hamiltonian in Eq.~\eqref{eq:potts}.
\subsubsection{Data generation}
To create a dataset the Wolff cluster algorithm~\cite{Wolff_89} is used to generate states corresponding to a thermal state with inverse-temperature $\beta = \ln(5)$. The first 1000 samples are discarded as burn-in and the data is thinned by a factor of 10. The training-set consists of 4000 shots and a validation set of 1000 shots, withheld from the training.

\subsubsection{Training details}
\begin{figure}[htp!]
    \centering
    \includegraphics[width=0.95\columnwidth]{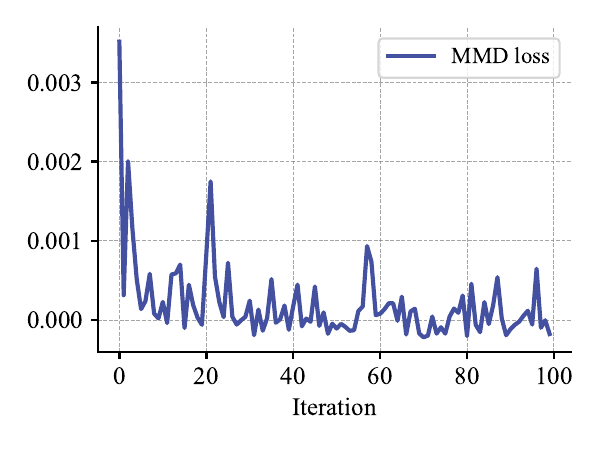}
    \caption{\textbf{Training loss for the Potts Model:} the MMD loss function in the IQP circuits training incorporating symmetry into the ansatz.}
    \label{fig:training_potts}
\end{figure}

The training is undertaken using the \textit{Adam} classical optimizer~\cite{kingma_2017} with a learning rate of 0.05. The number of operators sampled to evaluate the MMD was set to $N_\text{ops} = 4000$ and the number of uniformly sampled strings samples used to evaluate the expectation values was set to $N_\text{s} = 4000$. 
The bandwidth the MMD was chosen using Eq.~\eqref{eq:mu_bandwith}, with $\mu = \{2,6,12\}$. The trainable parameters were initialized by sampling from a normal distribution with zero mean and standard deviation $1/100$. The training results are shown in Fig.~\ref{fig:training_potts}, demonstrating that the model is able to learn.

\begin{figure}[htp!]
    \centering
    \begin{subfigure}[t]{0.48\textwidth}
        \centering
        \includegraphics[width=\linewidth]{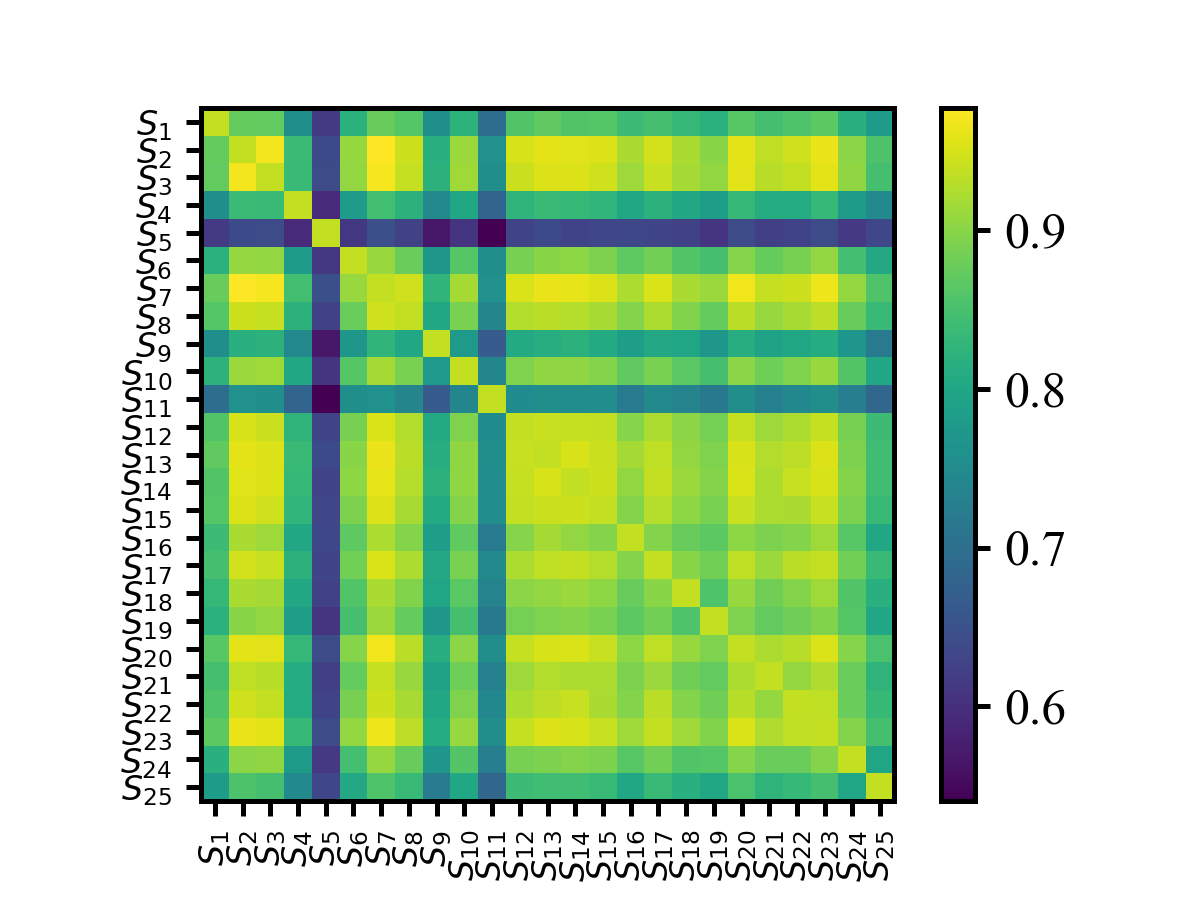}
        \caption{Generation data}
        \label{fig:potts_gen_corr}
    \end{subfigure}
    \begin{subfigure}[t]{0.48\textwidth}
        \centering
        \includegraphics[width=\linewidth]{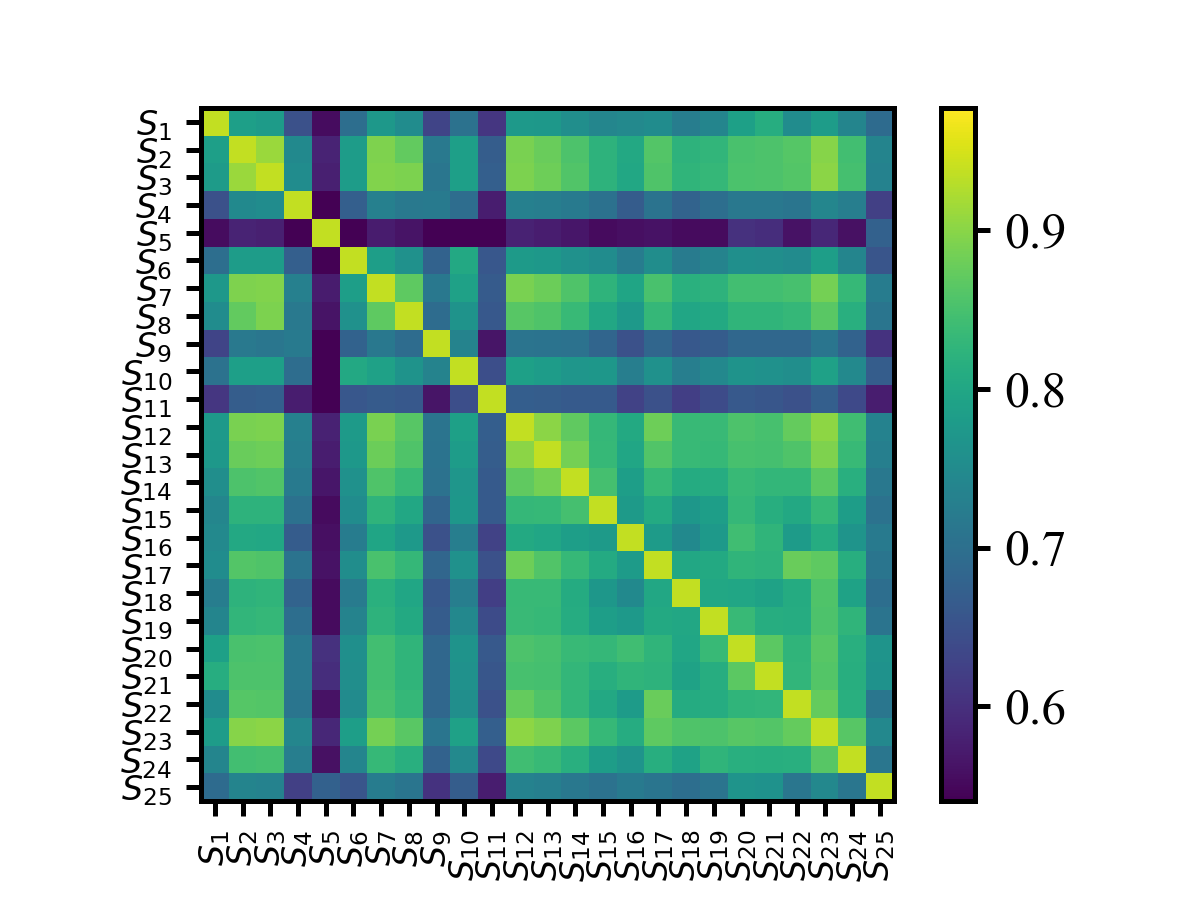}
        \caption{Validation data}
        \label{fig:potts_val_corr}
    \end{subfigure}
    \begin{subfigure}[t]{0.48\textwidth}
        \centering
        \includegraphics[width=\linewidth]{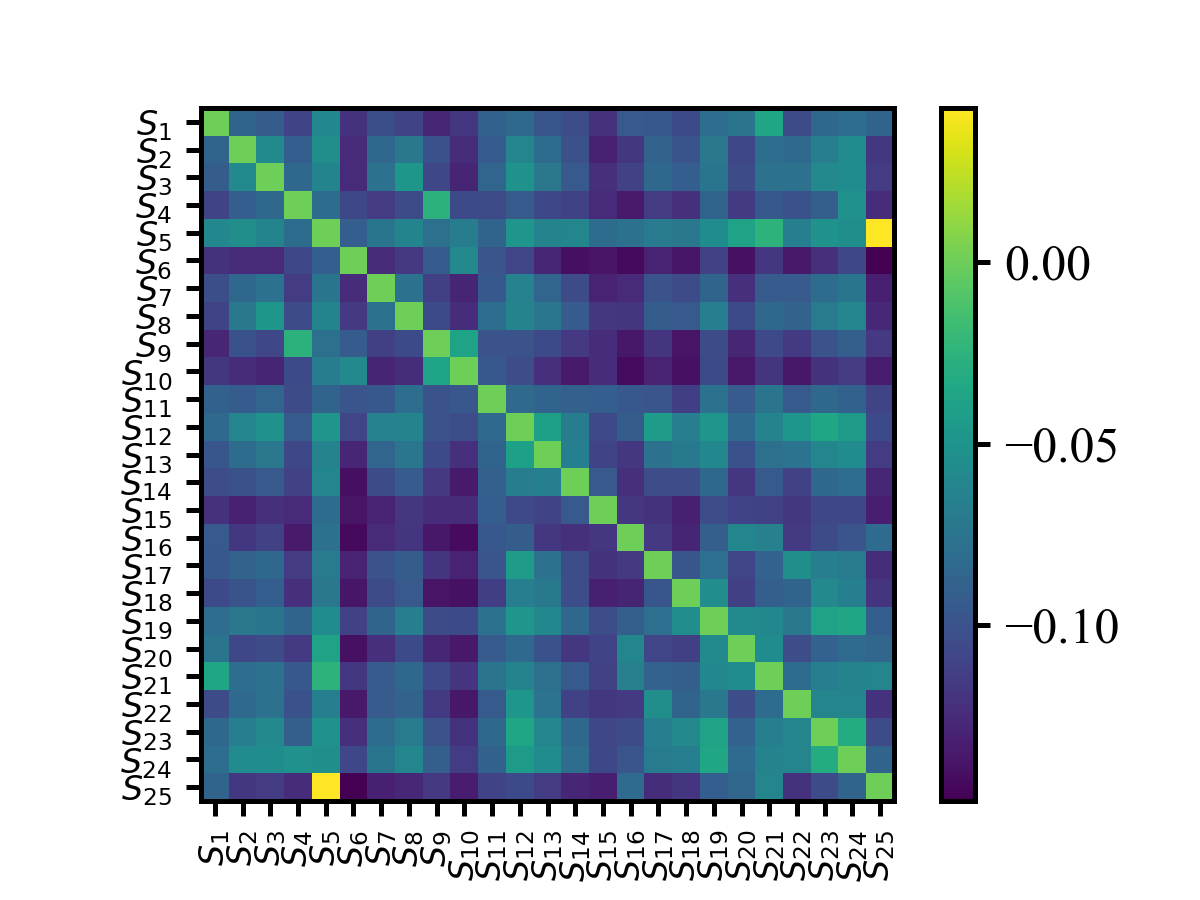}
        \caption{Validation data $-$ Generation data}
        \label{fig:potts_gen_val}
    \end{subfigure}
    \caption{\textbf{Correlation between spins in the Potts model}: (a) The estimated correlation from the IQP circuit. (b) The correlation in the validation data. In both panels, the matrices have been calculated with $\langle \mathbbm{1}(S_i=S_j)\rangle - 1/d$. (c) The difference between the generated correlation and the validation data. }
    \label{fig:potts_corr}
\end{figure}

\subsubsection{Validation}

Comparing the trained model to the withheld validation data gives a $\text{MMD}^2$ of $(8.1\pm 3.3) \times 10^{-5} $. Fig.~\ref{fig:potts_corr} shows $\langle \mathbbm{1}(S_i=S_j)\rangle - 1/d$ for the validation data as well as for the IQP circuit. One can see that the generated correlation is closer to the validation than the correlation considered in the main-text for the calorimeter data. This demonstrates the importance of exploiting symmetry in the data, where possible.

\section{Additional calorimeter results}
\label{app:add_results}
The main text focused on systems of size 6 (Sec.~\ref{sec:6_train},\ref{sec:6_val}) and 12 (Sec.~\ref{sec:12_train},\ref{sec:12_val}) pixels, for which the proposed approach demonstrated good performance. In this appendix, preliminary results for a larger system consisting of 25 pixels with qudit dimension $d=16$ (100 qubits required) is presented. While the model continues to capture the overall structure of the target distribution, the generated samples exhibit weaker correlations, highlighting a limitation of the current ansatz. Improving the reproduction of correlations is an important direction for future work and may benefit from more expressive circuit architectures, alternative parameterized gates, or loss functions that explicitly incorporate correlation information.
It should be emphasized that the present study does not constitute an extensive investigation of large-scale systems. The results were obtained using a limited number of operators $N_{\mathrm{ops}}$ and samples $N_\text{s}$, chosen to keep the computational cost manageable. Even under these constraints, training times were on the order of several hours. A more systematic exploration of larger models, increased sampling statistics is left for future work.

\subsubsection{Training details}

\begin{figure}[htp!]
    \centering
    \includegraphics[width=0.95\columnwidth]{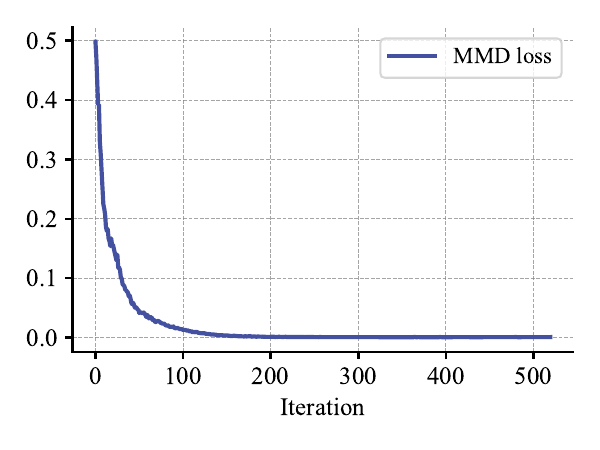}
    \caption{\textbf{Loss function for 25 pixels images and qudit dimension $\mathbf{d=16}$:} The MMD loss function in the IQP circuits training with kernel bandwidth $\sigma=7$ and scale for the initial gates parameters of $0.05$.}
    \label{fig:mmd_f25_d16}
\end{figure}

The model considered in this experiment contains a total of 5050 trainable parameters, initialized following Sec.~\ref{sec:iqp_training} with a scale factor of 0.05. Fig.~\ref{fig:mmd_f25_d16} shows the evolution of the MMD loss function during training of the IQP circuit, where the kernel bandwidth was set to $\sigma=7$. The observed decrease in the loss demonstrates that the model is able to learn relevant features of the target distribution despite the increased system size.
Due to the larger number of parameters and the increased computational cost associated with estimating the loss function, training was more demanding than in the smaller-system experiments presented in the main text.

\subsubsection{Validation}

\begin{figure}[htp!]
    \centering
    \includegraphics[width=1\columnwidth]{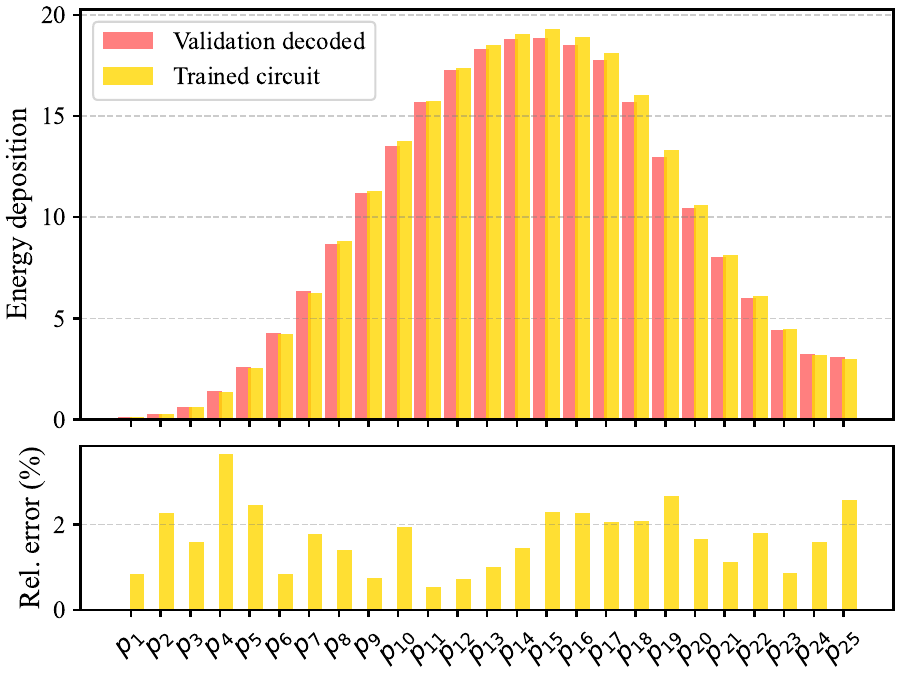}
    \caption{\textbf{Mean energy deposition for 25 pixels images with $\mathbf{d=16}$:} (\textit{top panel}) Energy deposition values of rescaled integer validation distribution (\textit{red}) and the expectation value of the number operator with trained IQP circuit (\textit{yellow}) and 1000 samples. (\textit{bottom panel}) The relative error of the means, expressed as a percentage ($\%$), depends on the number of samples considered.}
    \label{fig:samples_f25d16}
\end{figure}

The trained model is compared to the withheld validation data, giving a $\text{MMD}^2$ of $ (9.7 \pm 0.8) \times 10^{-4}$. 
Following the procedure outlined in the main text, the trained parameters are used to compute the expectation value of the number operator for each pixel, Eq.~\eqref{eq:number_op}. Fig.~\ref{fig:samples_f25d16} compares the results (\textit{yellow}) with the validation dataset (\textit{red}). The average energy deposition presents a relative error around $2\%$, showing good agreement with the validation data.

\begin{figure*}[ht!]
    \centering
    \begin{subfigure}[t]{0.49\textwidth}
        \centering
        \includegraphics[width=\linewidth]{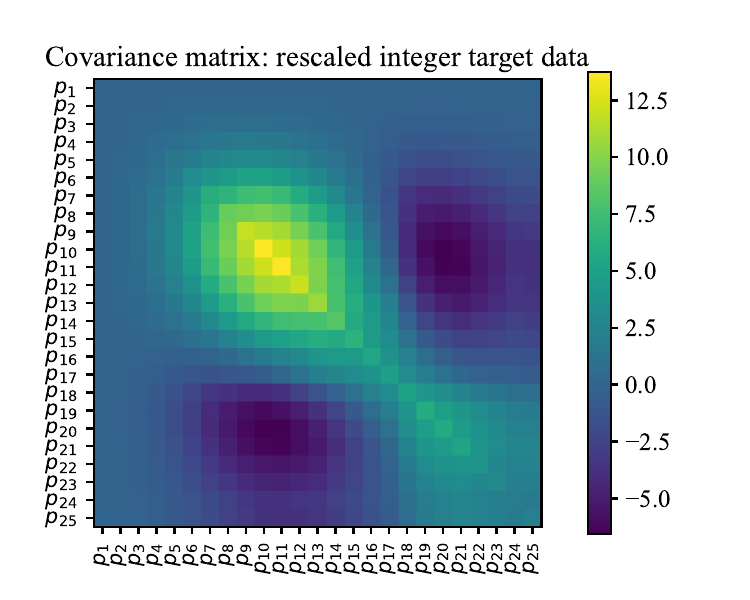}
        \caption{Covariance validation data}
        \label{fig:cov_f25_d16_target}
    \end{subfigure}
    \hfill
    \begin{subfigure}[t]{0.49\textwidth}
        \centering
        \includegraphics[width=\linewidth]{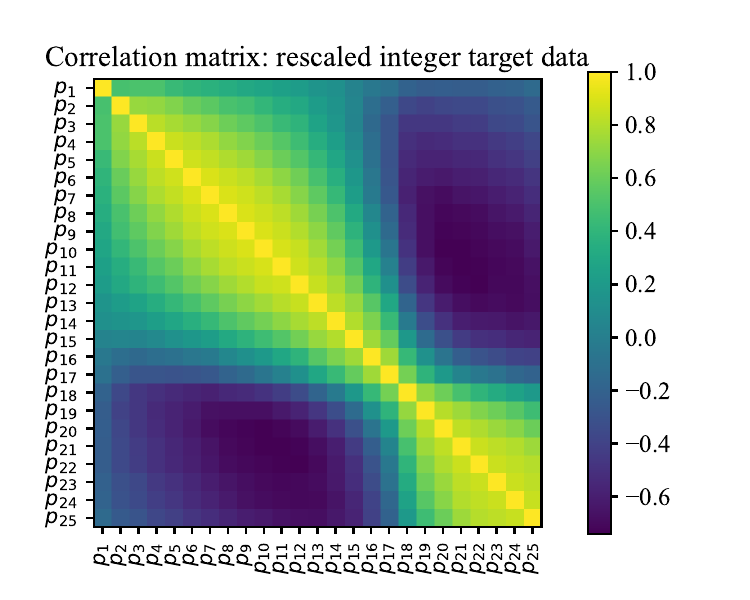}
        \caption{Correlation validation data}
        \label{fig:corr_f25_d16_target}
    \end{subfigure}

    \begin{subfigure}[t]{0.49\textwidth}
        \centering
        \includegraphics[width=\linewidth]{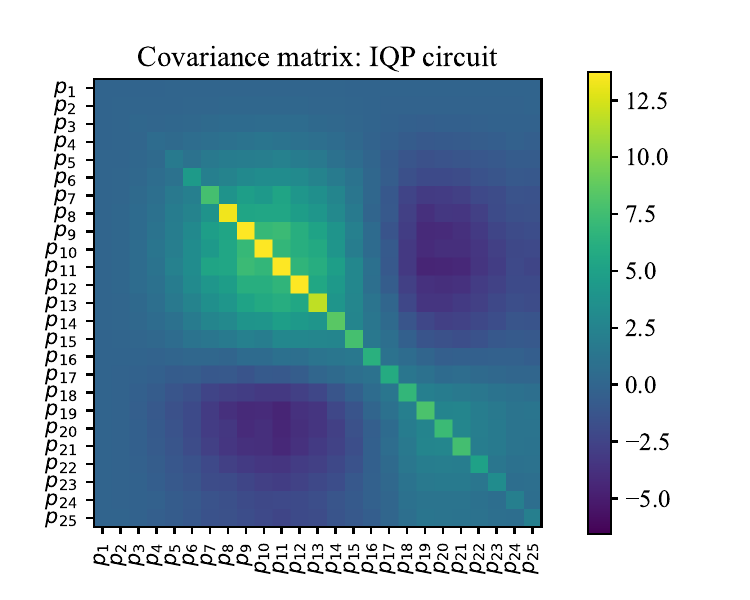}
        \caption{Covariance generation data}
        \label{fig:cov_f25_d16_iqp}
    \end{subfigure}
    \hfill
    \begin{subfigure}[t]{0.49\textwidth}
        \centering
        \includegraphics[width=\linewidth]{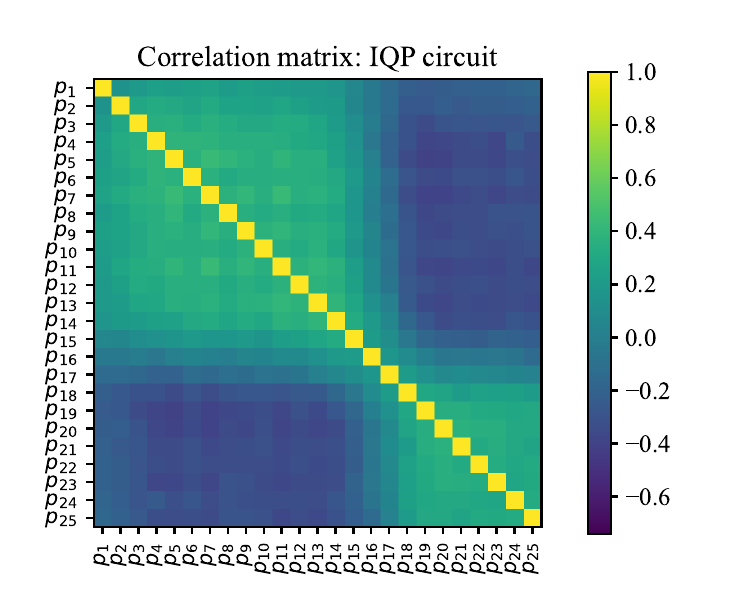}
        \caption{Correlation generation data}
        \label{fig:corr_f25_d16_iqp}
    \end{subfigure}
    \caption{\textbf{Covariance and correlation matrix for 25 pixels images and qudit dimension $\mathbf{d=16}$ :}  (a) Covariance matrix between pixels of the validation dataset. (b) Correlation matrix between pixels of the same set. (c) Estimated covariance matrix from the IQP circuit, obtained with 1000 samples. (d) Estimated correlation from the IQP circuit, obtained from the covariance calculation. The color scales of the bottom panels are the same as in the top panels.}
    \label{fig:cov_corr_f25_d16}
\end{figure*}
Figs.~\ref{fig:cov_f25_d16_target} and~\ref{fig:corr_f25_d16_target} depict the covariance and correlation matrices of the target validation dataset, while Figs.~\ref{fig:cov_f25_d16_iqp},~\ref{fig:corr_f25_d16_iqp} show the covariance and correlation matrices obtained from the trained IQP circuits. In the correlation results, the residual element-wise error is on average $|\Lambda| = 0.251$, while the off-diagonal correlation coefficient is $r=0.968$. 

Improving correlation fidelity at larger system sizes is an important direction for future work. Potential avenues include modifying the loss function to better capture higher-order dependencies and exploring alternative parameterizations of quantum gates that may improve expressivity. We remark that the current training procedure requires on the order of hours per experiment, and extensive optimization for larger-scale settings has not yet been conducted.
A further limitation arises from the restricted number of optimization steps $N_{\mathrm{ops}}$ and sampled data points $N_\text{s}$, which constrain both performance and statistical accuracy of the learned distribution.

\section{Prior work on quantum machine learning applied to calorimeter images}
\label{app:qml_works}
In this appendix, a summary of some of the previous approaches to applying quantum machine-learning to generate calorimeter images is detailed. The focus is on methods that utilize parameterized quantum circuits, as opposed to prior work utilizing quantum annealers~\cite{tol2024, Hoque2024}. The aim is to examine previous approaches, discussing how the data has been encoded and the eventual limitations of the models. In this section, $X, Y,$ and $Z$ refer to the standard qubit Pauli-operators. 

\subsection{Quantum Circuit Born Machines}

Rudolph \textit{et al.}~\cite{rudolph2024trainability} consider $4 \times 4$ black-and-white calorimeter images. Their implementation consists of a Quantum Circuit Born Machine (QCBM) \cite{Cheng_2018, benedetti2019generative, coyle2020born}, which encodes the probability into the amplitudes of a state vector such that
\begin{equation}
    q_{\boldsymbol{\theta}}(\mathbf{x}) = \abs{ \bra{\mathbf{x}} U(\boldsymbol{\theta}) \ket{\mathbf{0}}}^2,
\end{equation}
where $U(\boldsymbol{\theta})$ is the unitary associated with a quantum circuit parameterized by $U(\boldsymbol{\theta})$. The IQP circuit considered in this paper is a specific example of a QCBM. The authors demonstrate the advantage of implicit loss-functions, i.e. loss-functions dependent on an average of samples taken from a distribution, over explicit loss-functions, i.e. loss-functions dependent on obtaining probabilities. The MMD is an example of the latter. The authors also introduce the MMD as a probabilistic mixture of observables, central to the train-on-classical approach explored in Ref.~\cite{recio2025train}. The direct numerical experiment on HEP data in this paper is limited by the requirement to perform numerical simulations. In this work the calorimeter images are converted to binary data.

\subsection{Amplitude encoding}

Encoding the intensity of each pixel as the amplitude of a quantum state was considered by Chang \textit{et al.} in Ref.~\cite{Chang_2021}. The model they consider is an example of a Quantum Generative Adversarial Networks (QGAN)~\cite{Dallaire_Demers_2018, Zoufal_2019}, where amplitude encoding means that the number of qubits required is logarithmic in the number of pixels. However, an exponential number of shots, in terms of the number of qubits, is then required to reconstruct the image classically. This is a major limitation of the approach. The authors suggest that this could be overcome by performing the desired image analysis directly on the quantum computer.

\subsection{The Quantum Angle Generator}

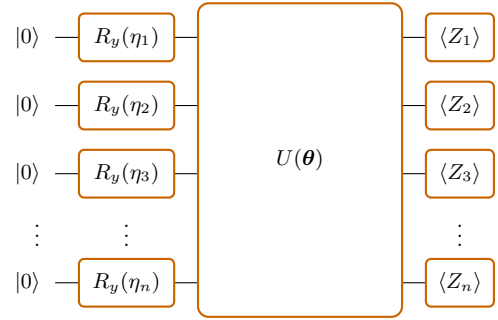
\begin{figure}[htp]
    \centering
\begin{tikzpicture}[scale=0.9, every node/.style={scale=0.9}]

\def\ywa{0}      
\def\ywb{-1.0}   
\def\ywc{-2.0}   
\def\ygap{-2.8}  
\def\ywd{-3.6}   

\def\xKr{1.05}   
\def\xW{1.15}    
\def\xRyl{2.2}   
\def\xBl{3.25}   
\def\xBr{6.25}   
\def\xMl{7.10}   
\def\xBrace{0.45}

\foreach \y in {\ywa,\ywb,\ywc,\ywd}
  {\draw (\xW,\y)--(\xMl-0.5,\y);}

\foreach \y in {\ywa,\ywb,\ywc,\ywd}
  {\node[font=\small,anchor=east] at (\xKr,\y){$\ket{0}$};}

\node[font=\small,anchor=east] at (\xKr,\ygap){$\vdots$};
\node[font=\small] at (\xRyl,\ygap){$\vdots$};
\node[font=\small] at (\xMl,\ygap){$\vdots$};

\draw[darkorange,thick,fill=white,rounded corners=4pt]
  (\xBl,\ywd-0.5) rectangle (\xBr,\ywa+0.5);
\node[font=\small] at ({(\xBl+\xBr)/2},{(\ywa+\ywd)/2})
  {$U(\boldsymbol{\theta})$};

\foreach \y/\sub in {\ywa/{\eta_1},\ywb/{\eta_2},\ywc/{\eta_3},\ywd/{\eta_n}}{
  \draw[darkorange,thick,fill=white,rounded corners=2pt]
    (\xRyl-0.7,\y-0.35) rectangle (\xRyl+0.7,\y+0.35);
  \node[font=\small] at (\xRyl,\y){$R_y(\sub)$};
}

\foreach \y/\sub in {\ywa/{1},\ywb/{2},\ywc/{3},\ywd/{n}}{
  \draw[darkorange,thick,fill=white,rounded corners=2pt]
    (\xMl-0.5,\y-0.35) rectangle (\xMl+0.5,\y+0.35);
  \node[font=\small] at (\xMl,\y){$\langle Z_{\sub}\rangle$};
}

\end{tikzpicture}
\caption{\textbf{QAG circuit structure:} A first layer of noise-loading $R_y$ rotations is followed by a trainable unitary $U(\boldsymbol{\theta})$, where $\boldsymbol{\theta}$ denotes the trainable parameters. The last layer denotes the estimation of the Pauli-Z expectation values for each qubit.
}
    \label{fig:q_circuit_qag}
\end{figure}

The Quantum Angle Generator (QAG), applied to calorimeter data in Ref.~\cite{rehm2024precise}, encodes $n$ features into $n$ qubits. Each grey-scale pixel is encoded as the $Z$ expectation value of the corresponding qubit. The circuit consists of a trainable unitary $U(\boldsymbol{\theta})$, where $\boldsymbol{\theta}$ denotes the trainable parameters. To allow for the generation of multiple images, an $n$ dimensional latent variable, denoted by $\boldsymbol{\eta}$, is sampled from a probability distribution. Each component of $\boldsymbol{\eta}$ is assumed to have been drawn from an identical and independent distribution. A $Y$-rotation is used to load each component of $\boldsymbol{\eta}$ onto the corresponding qubit, prior to the trainable unitary. The QAG model is illustrated in Fig.~\ref{fig:q_circuit_qag}. 
The work of Rehm \textit{et al.}~\cite{rehm2024precise} investigate the $n=8$ case. 

A notable challenge for the QAG is making sure that the latent variable $\boldsymbol{\eta}$ is drawn from an appropriate distribution. The rest of this section illustrates the role $\boldsymbol{\eta}$ plays in the average value output by the QAG model. Let $i$ denote the $i^{\text{th}}$ qubit, with $i$ between $1$ and $n$. At the end of the circuit $\langle Z_i \rangle$ is measured to give the outcome of pixel $i$. The average value of $\langle Z_i \rangle$ is given by
\begin{equation}
    \text{Avg}(\langle Z_i \rangle) = \mathbb{E}_{\boldsymbol{\eta}}  \left[ \text{Tr}\left(U^\dagger (\theta) Z_i U(\theta) \bigotimes_{k=1}^n \rho_0(\eta_k) \right)\right],
\end{equation}
where
\begin{equation}
    \rho_0(\eta_k) = R_y(\eta_k)\ket{0}\bra{0} R_y(-\eta_k)
\end{equation}
and $\eta_k$ is the $k^{\text{th}}$ component of $\boldsymbol{\eta}$ and $R_y$ denotes a single qubit $Y$ rotation. It follows that
\begin{align}
    \text{Avg}(\langle Z_i \rangle) = & \text{Tr}\left(U^\dagger (\theta) Z_i U(\theta)  \mathbb{E}_{\boldsymbol{\eta}}  \left[ \bigotimes_{k=1}^n \rho_0(\eta_k) \right]\right)\nonumber\\
    = & \text{Tr}\left(U^\dagger (\theta) Z_i U(\theta)  \bigotimes_{k=1}^n \mathbb{E}_{\eta_k}  \left[ \rho_0(\eta_k) \right]\right)\nonumber\\
    = & \text{Tr}\left(U^\dagger (\theta) Z_i U(\theta)  \bigotimes_{k=1}^n \overline{\rho}\right),
\end{align}
where $\overline{\rho} = \mathbb{E}_{\eta}  \left[ \rho_0(\eta) \right]$, exploiting the fact each $\eta_k$ is assumed to be drawn from an identical and independent distribution.

Let $\vec{\lambda} = (\lambda_0, \lambda_1, \dots, \lambda_{2^n-1})$ be the ordered eigenvalues of $\overline{\rho}^{\otimes n}$ such that $\lambda_0 \geq \lambda_1 \geq \dots \geq \lambda_{2^n-1}$. From the Schur–Horn theorem \cite{watrous2018theory} the unitary $U(\theta)$ can only map the initial density operator to a new density operator with ordered diagonal values $\vec{p} = (p_0, p_1, \dots, p_{2^n-1})$ (again with $p_0 \geq p_1 \geq \dots \geq p_{2^n-1}$) if $\vec{\lambda}$ majorizes $\vec{p}$. This can be used to bound $\text{Avg}(\langle Z_i \rangle)$. Denoting each bitstring by $x$ which occurs with probability $p(x)$, and the $i^{\text{th}}$ element of $x$ by $x_i$, then:
\begin{align}
    \text{Avg}(\langle Z_i \rangle) & = \sum_{x: x_i=0} p(x) - \sum_{x: x_i=1} p(x) \\
    & = 2 \sum_{x: x_i=0} p(x) - 1\\
    & \leq 2 \sum_{k=0}^{2^{n-1}-1} \lambda_k - 1.
\end{align}
where in the last line we have used that exactly half of the bit strings have $0$ in the $i^{\text{th}}$ position plus the majorization condition. The lower-bound follows similarly to give
\begin{equation}
    \abs{\text{Avg}(\langle Z_i \rangle)} \leq 2 \sum_{k=0}^{2^{n-1}-1} \lambda_k - 1.
\end{equation}
Taking as an illustrative example:  $\boldsymbol{\eta}$ drawn from a uniform distribution between $0$ and $2\pi$, then $\overline{\rho} = I/2$ and $\abs{\text{Avg}(\langle Z_i \rangle)} \leq 0$ (i.e. equals zero). This shows that the choice of $\boldsymbol{\eta}$ can change the reachability of the QAG, requiring careful selection of $\boldsymbol{\eta}$ (or rescaling of the data).

\subsection{Quantum Feature Amplification Network}

In a recent work a model referred to as a Quantum Feature Amplification Network (QFAN) was proposed by Slim \textit{et al.}~\cite{slim2026_qfan} to tackle calorimeter data. The paper decouples the number of features, using three qubits to learn images of size 12 and 25 pixels. Table 1 in Ref.~\cite{slim2026_qfan} implies 10 qubits are sufficient for grey-scale images with 40,500 pixels. After the initial training with three qubits, a subsequent classical training step is employed to align the quantum-generated outputs with the target training data.

\subsection{Sparse binary IQP circuits}
\label{app:iqp_sparse}
During the preparation of this manuscript, a paper applying binary IQP circuits to the calorimeter data was published by Slim \textit{et al.} \cite{slim2026iqpbornmachinecalorimeter}. The IQP circuits considered in the work are sparse in terms of two qubit gates between qubits. The diagonal part of the IQP circuits considered in Ref.~\cite{slim2026iqpbornmachinecalorimeter} consist of single qubit rotations and two-qubit rotations defined on a graph $G=(V,E)$.  This graph $G$ is sparse in the sense that the average degree is fixed as $n = \abs{V}$ is increased. The sparse nature of the graph limits the possible correlations achievable by the model. This can be seen as follows:
\begin{enumerate}
    \item If two qubits in a binary IQP circuit share neither common nodes nor an edge, the two qubits have zero covariance (hence zero correlation).
    \item For the fixed average degree graphs considered in Ref.~\cite{slim2026iqpbornmachinecalorimeter}, the fraction of nodes with zero correlation approaches 1 as $n$ approaches infinity.
\end{enumerate}
The rest of this section elaborates on these points.

Consider the two point function $\langle  Z_i Z_j \rangle$ for a binary IQP circuit with $n$ qubits.  Since all diagonal gates commute, the Heisenberg evolution of any $Z$ only has support on adjacent qubits in $G$ for binary IQP circuits. Therefore if nodes $i$ and $j$ have no common support, then
\begin{equation}
    \langle  Z_i Z_j \rangle = \langle  Z_i \rangle \langle Z_j \rangle,
\end{equation}
i.e. there is no covariance between qubits corresponding to nodes with neither a shared edge nor qubits adjacent to both $i$ and $j$. If the covariance is $0$ the correlation is zero.

The IQP circuits in Ref.~\cite{slim2026iqpbornmachinecalorimeter} are defined on Erd\H{o}s--R\'enyi graphs with fixed average degree $\Delta$ and $n$ nodes. That is to say, each edge in the graph is selected with probability of $p = \Delta/(n-1)$. The probability that two nodes do not share an edge is $1-p$. The probability that given two nodes, they are both connected to a third node is $p^2$. Therefore the probability that two nodes $i$ and $j$ neither share an edge nor are adjacent to any of the same nodes is:
\begin{multline}
    P(\text{nbr}(i) \cap \text{nbr}(j) = \emptyset ) \\ = \left(1-\frac{\Delta}{n-1}\right)\left(1-\left(\frac{\Delta}{n-1}\right)^2\right)^{n-2},
\end{multline}
where nbr$(k)$ denotes the set of nodes adjacent to node $k$. As $n$ tends to infinity, $P(\text{nbr}(i) \cap \text{nbr}(j) = \emptyset )$ tends to 1. If $\text{nbr}(i) \cap \text{nbr}(j) = \emptyset $ then 
\begin{equation}
    \text{Cov}(Z_i, Z_j) = \langle Z_i Z_j\rangle - \langle Z_i \rangle \langle Z_j \rangle  = 0.
\end{equation}
So the number of qubits with non-zero correlation vanishes as $n$ tends to infinity. At $\Delta = 6$ and $n=64$ considered in Ref.~\cite{slim2026iqpbornmachinecalorimeter}, $P(\text{nbr}(i) \cap \text{nbr}(j) = \emptyset ) \approx 0.514$. Hence, for the average Erd\H{o}s--R\'enyi graph with $\Delta= 6$, $n=64$, approximately half of the qubit correlations should be zero. This suggests a possible limitation of using sparse-IQP circuits.

Beyond the choice of sparse binary IQP circuits, Slim \textit{et al.}\ train directly on the correlation of the model. They also train over several copies of the model, averaging the loss function over each copy. Since the IQP circuit remains unchanged between copies, the drawbacks of sparse IQP circuits are still present.

\bibliography{bibliography}

\end{document}